\documentclass[epsfig,usegraphicx,useAMS,usenatbib]{mn2e}
\usepackage{rotating}

\title[The GALEX-SDSS $NUV$ \& $FUV$ Flux Density and Local Star-Formation Rate]{The GALEX-SDSS $NUV$ \& $FUV$ Flux Density and Local Star-Formation Rate}
\author[A. S. G. Robotham]{A. S. G. Robotham$^{1,2}$\thanks{asgr@st-and.ac.uk} and S. P. Driver$^{1,2}$\thanks{spd3@st-and.ac.uk}\\
$^{1}$Scottish Universities Physics Alliance, SUPA \\
$^{2}$School of Physics and Astronomy, University of St Andrews, KY16 9SS, UK}
\begin{document}

\date{}

\pagerange{\pageref{firstpage}--\pageref{lastpage}} \pubyear{2010}

\maketitle

\label{firstpage}

\begin{abstract}
We calculate the local UV flux density in the GALEX MIS $FUV$ and $NUV$ bands using redshifts provided by SDSS DR7. Luminosity functions are calculated for the overlapping MIS and SDSS sample, allowing flux densities to be measured and the local star formation rate (SFR) to be calculated using volumes much larger than previous $FUV$ based estimates. We calculate flux densities for a dust corrected low redshift ($0.013 \le z \le 0.1$) sample of $f_{\nu-FUV}=22.24 \pm 3.13 $ $\times 10^{25} h$ ergs s$^{-1}$ Hz$^{-1}$ Mpc$^{-3}$, $f_{\nu-NUV}=38.54 \pm 5.30 $ $\times 10^{25} h$ ergs s$^{-1}$ Hz$^{-1}$ Mpc$^{-3}$. The star formation rate density found is $0.0312\pm0.0045$ $h$ M$_{\odot}$yr$^{-1}$ Mpc$^{-3}$. This is larger than published rates recently found using the UV implied SFR, though the major discrepancy is the correction made for dust attenuation and once this is dealt with consistently the results agree well. These values are also consistent with recent H$_{\alpha}$ derived SFRs. Once cosmic variance is taken into account most of the recent SFRs at low redshift ($z \le 0.3$) found in the literature can be brought into agreement, however the lowest redshift values ($z \le 0.045$) do appear to be significantly lower.
\end{abstract}


\section{Introduction}

In its simplest form the UV to near-IR output of the entire Universe can be predicted with remarkably few prescriptions. Primarily we need to know the initial mass function (IMF) for stars, be this constant \citep[e.g.][ and references therein]{Wyse1997} or varying \citep[e.g.][]{Wilkins2008}; the evolution of stars once they have formed \citep[e.g.][]{Bruzual2003,Maraston2005} and the rate at which baryons have been turned into stars throughout the history of the Universe to our current epoch \citep{Hopkins2006}. On top of this we must understand the role of dust in attenuating the light pumped into the cosmos \citep[e.g.][]{Calzetti2000,Tuffs2004,Driver2008}. Though we can state this quite succinctly, each component of the puzzle is challenging for both theory and observation. In our ideal picture we would be able to state that the Universe produces stars at a rate determined from the star formation history (SFH) and with a mass distribution given by the IMF. To predict the local Cosmic Energy Distribution (CSED), all we need to do is model the evolution of stars using a population synthesis model and account for the role of dust in absorbing and re-emitting the light in the mid and far-IR.

In practice, our best current observations and theoretical models cannot perfectly measure or predict any of these separate components, and thus the $z=0$ light output of the Universe is non-trivial to explain. In terms of predictions, weak constraints on IMFs have been made through high resolution modelling \citep{Bonnell2001}. Equally, population synthesis models have become increasingly sophisticated, but no purely theoretical models have created a star as we observe them, and major issues exist with the latter stages of stellar evolution where dredge up and mass loss have a significant effect on the radiation output of stars but are difficult to model \citep{Marigo2008}. The star formation rate (SFR) has been predicted in a variety of simulations \citep[e.g.][]{Cole2000,Hernquist2003,Nagamine2006} but these are ultimately limited by our observations (because models are designed to fit the data) and the assumed IMF--- clearly the conversion between UV or H$_{\alpha}$ output and SFR can differ between reality and simulation. These theoretical difficulties highlight the full complexity of the problem, and ultimately mean that observations are still the leading light in building our understanding of the flux output of the Universe.

Observationally the IMF can be measured to some extent in star forming regions. This is difficult to do directly since the most massive stars are short-lived and have a delayed proto-stellar period compared to less massive stars. This short window of existence makes it unlikely that they will be observed directly, placing significant uncertainty on the high mass slope and turnover of the IMF. Star forming regions are also expected to be highly obscured by dust and much star forming activity will be hidden entirely. Despite these restrictions the IMF has been measured directly in various regions of the Milky Way \citep{Scalo1998}, with the general finding that it is not radically dissimilar to a Salpeter IMF \citep{Salpeter1955}. Observations also play a key role in producing stellar models since spectral measurements of stars are used as templates in synthesis codes to evolve stars and galaxies. A vast amount of effort has been invested into measuring the cosmic SFH, surveys have been conducted at a large range of redshifts \citep[see compendiums by][and references therein]{Wilkins2008,Hopkins2006} utilising different tracers in order to produce a SFR density as a function of look-back time, creating the, now ubiquitous, Lilly-Madau diagram \citep{Lilly1996,Madau1996,Madau1998}. Despite these tremendous efforts, the SFH of the Universe is not well constrained. In particular, the significant effects of cosmic variance (see Driver \& Robotham 2010) and even determining the true volume of space surveyed at high redshift makes constraining the SFR at early epochs extremely difficult. Current results allow the cosmic SFR to be roughly constant for $z>2$ or even declining sharply, the single consensus is that it was not any higher in the earliest epochs of the Universe. As well as being a key ingredient in calculating the CSED, some form of IMF is required to convert all star formation tracers into conventional units of SFR. Often this conversion is done implicitly, but there is enough convergence in different IMF models that the resultant SFRs are not hugely sensitive to the IMF even for different tracers.  

The vast majority of UV output in the $FUV$ and $NUV$ bands can be attributed to high mass and short-lived stars. As such any evolution in the SFR can be probed quite precisely, providing a suitably short redshift (i.e.\ temporal) baseline is chosen. Since large volumes are required to ensure meaningful sampling statistics, large coverage areas are needed also--- especially true for probing the local Universe. The local implied SFR is an extremely important normalisation point for models that attempt to predict the cosmic star formation history (SFH), providing a value that is effectively independent of all but the most recent star formation. Since the local SFR can be measured with the greatest accuracy, it provides the main constraint to the normalisation of SFH.

Much recent work has attempted to measure the local SFR, predominantly through calculating and integrating out the UV \citep{Wyder2005,Budavari2005,Salim2007,Wyder2007}, H$_{\alpha}$ \citep{PerezGonzalez2003} or OII \citep{Gallego2002} luminosity functions. However since these studies were published a huge amount of data has been released, increasing the number of galaxies with both measured UV fluxes and redshifts. Considering purely $FUV$ derived SFRs the increase is ~15$\times$ \citep{Wyder2005}, although more recent work has used $FUV$ data as part of a fitted SED derived SFR, making the data increase in this work slightly more than 1.3$\times$ \citep{Salim2007}. This work calculates local flux densities in the redshift ranges $0.013 \le z \le 0.1$ for the GALEX $FUV$ and $NUV$ bands, and calculates the local SFR based on the $FUV$ flux density. Throughout we use a cosmology of H$_{0}=$100kms$^{-1}$, $\Omega_{M}=0.3$ and $\Omega_{\Lambda}=0.7$. The work was carried out using a combination of publicly available GALEX MIS survey data \citep{Morrissey2007} and SDSS DR7 redshift data \citep{Abazajian2009}. The GALEX MIS data are available from the MAST website\footnote{http://galex.stsci.edu/GR4/}, and the SDSS DR7 is available from the CasJobs SQL website\footnote{http://casjobs.sdss.org}. Since the $FUV$ detecter on GALEX is no longer operational, the sample presented here is close to the largest that will be available in the near to mid future.

\section{GALEX-SDSS Data Processing}

\subsection{Data Acquisition}

\begin{figure*}
\begin{center}
\includegraphics[width=5in]{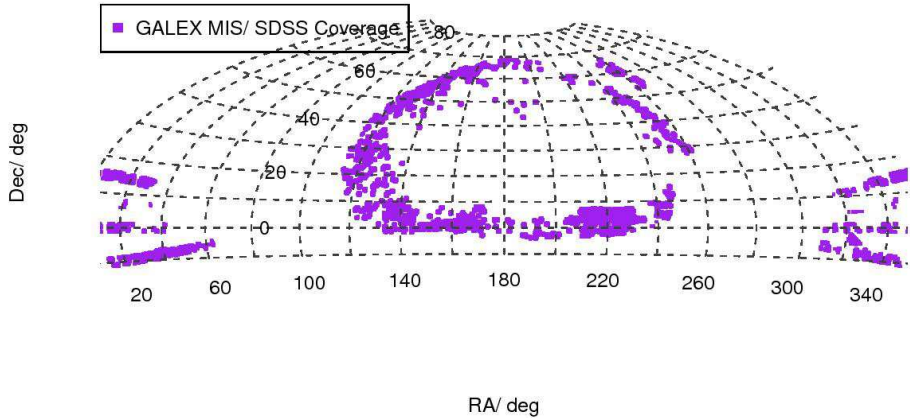}
\caption{\small  Plot showing the overlap between GALEX MIS and SDSS DR7 redshifts. The total shared area is 833.13 sq. deg.}
\label{GALEXcover}
\end{center}
\end{figure*}

To generate our NUV and FUV luminosity function we require all of the potential object matches between the GALEX MIS and SDSS DR7. The first step was downloading the full MIS catalogue for tiles with exposure times greater than 1,400 seconds and $E(B-V)<0.08$, ensuring uniformly adequate depth and minimal Galactic dust corrections. To extract the overlap search region required for this work cone searches were made through {\bf Topcat} \citep{Taylor2005}, using the MIS tile centres and a search radius of $0.6^{\circ}$. Doing searches with {\bf Topcat} proved to be a more efficient technique than using the CasJobs service since the MIS tiles were spread out over such a complex survey geometry. We further cut this sample at an SDSS $r_{\rm petro}<20.5$mag, leaving a superset of ~15M potential GALEX objects in the SDSS catalogue, and ~6.4M primary photometry GALEX objects in the full MIS catalogue. Figure \ref{GALEXcover} shows the complex overlapping coverage between the GALEX MIS and SDSS surveys.

Due to the complex overlap between the GALEX MIS and SDSS DR7 surveys, calculating the area coverage is non trivial. To estimate the area we pixelated the common region of a cylindrical projection and counted the number of joint GALEX-SDSS galaxies inside each pixel. Those pixels that had 8 immediate neighbours with 1 or more counts were considered to have complete coverage. The total number of objects contained within these `complete' pixels was compared to the total number of objects over all pixels, and the total area of the `complete' pixels was scaled by the same ratio. Whilst this process is sensitive to the pixel size chosen, there is a steady plateau of values for pixel sizes where the ratio between the number of `complete' pixels and the number of partial coverage pixels (1 or more counts) is highest. The pixel size where this statistic is highest should give the best estimate of the area, since the correction is smallest. The peak found for the GALEX-SDSS overlap data was for a pixel of size $15^{\prime}\times15^{\prime}$ at the equator, which gives 36.4\% `complete' pixels. When the number count scaling is applied the total common GALEX-SDSS area is found to be 833.13 deg$^{2}$.

To assess errors associated with our area estimate, simulations were made of survey overlaps much like that used in this work--- patches of contiguous overlapping GALEX size fields using the same number densities as found empirically. These were generated in areas ranging from single GALEX field scales to 1000+ deg$^{2}$, with a restriction that no contiguous block of GALEX fields can cover more than 100 deg$^{2}$-- this is to ensure we investigate the worst case scenario of extremely patchy overlap. The results displayed marginal biasing towards overestimating the area, but errors are dominated by the noise in the estimation. The standard deviation in the area scatter is $\sim$2.30\% for areas below 50 deg$^{2}$, and drops to ~0.34\% for area between 800 and 900 deg$^{2}$. This implies an area estimate of $833.13 \pm 2.83$ deg$^{2}$. In practice the estimation should be better than this because the true GALEX-SDSS overlap contains much larger contiguous regions than allowed in the simulations, and the errors drop much faster for such structures than those implied by adding small areas and calculating the errors in quadrature.

\subsection{GALEX Flux Correction}

GALEX typically has FWHM imaging of ~5$^{\prime\prime}$, much greater than the ~1.5$^{\prime\prime}$ experienced with SDSS which we use for redshifts. To combat potential ambiguity in GALEX flux assignment a novel approach has been used based on the fact that the object centroid available for GALEX data are a flux moment based measurement. This means the centroid of the system is defined as the centre-of-mass (COM) in flux space.

The COM in any system can be calculated by finding the COM of subsystems in order to simplify the problem, e.g.\ in the Earth--Moon system we can simplify the Earth and Moon as point sources and calculate the complete system COM accordingly. The same is true for calculating centroids in an image: in a symmetrical extended source (e.g.\ galaxy) the COM is at the peak of the light profile, and for a point source (e.g.\ star) it is also at the peak. If these two objects were incorrectly merged and the COM of the system was found to be half way between these two points then we would know that they each possess an equal amount of flux. For this work, if we only had access to the merged GALEX data then the problem would be intractable, but the associated SDSS data contains the centroid locations of all the likely component objects that were merged in GALEX.

Since the object detection threshold used for GALEX will also be affected by significant de-blending issues when sources are close, we make a match for all object centroids (stars and galaxies) within the Petrosian r$_{90}$ distance (the radius which encloses 90\% of the $r$-band flux) for primary GALEX sources (this flag selects the best GALEX photometry in overlapping regions) and tag these as objects that potentially contribute some of the $NUV$/$FUV$ flux from GALEX. Of the potential SDSS objects, 273k match ambiguously and ~2.23M match unambiguously, so ~83\% of SDSS sources down to $r_{\rm petro}<20.5$mag have a GALEX object associated with it, of which ~74\% have an unambiguous 1--1 source match. At this stage the matching explicitly includes stars--- this is important since often stars are the contributor to the UV flux and must be kept in the catalogue until after the flux has been redistributed.

\begin{figure}
\begin{center}
\includegraphics[width=3in]{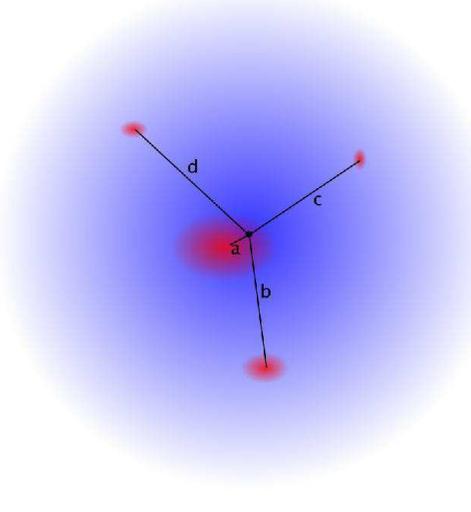}
\caption{\small  Schematic showing the sort of ambiguous matches that occur between SDSS (red resolved galaxies) and GALEX (large blue unresolved detection). In practice such ambiguity is rare- the majority of matches are unambiguous, and there are only a few dozen GALEX sources with 4 or more SDSS matches.}
\label{GALEXflux}
\end{center}
\end{figure}

The flux for ambiguous targets is subsequently distributed according to a few basic rules. Firstly, the SDSS object that matches closest to the GALEX centroid is designated the primary match, and if no other matching object is within 2.5 magnitudes of the $g$-band flux then all of the GALEX flux is assigned to the primary target (this proved to be a necessary addition to prevent large objects from being shredded by spurious near-by SDSS detections). For this decision the $g$-band was chosen due to its significant photometric fidelity compared to the $u$-band data. Assuming some brighter sources remain as potential matches the next phase is to distribute flux according to the linear separation between the GALEX and SDSS centroids, e.g.

\begin{eqnarray}
A&=&\displaystyle\sum_{i=1}^{n} \frac{1}{r_{i}} \\
f_{i}&=&\frac{f_{T}}{r_{i}A}
\end{eqnarray}

\noindent where  $r_{i}$ is the separation between the GALEX source centre and the $i^{th}$ SDSS object, $f_{T}$ is the total GALEX flux, $n$ is the number of SDSS objects that meet the matching criteria specified above and $f_{i}$ is the implied GALEX flux for the $i^{th}$ SDSS object.

Figure \ref{GALEXflux} is a schematic of how the GALEX UV flux (the extended blue source) can be associated with multiple SDSS sources (the smaller red objects). The source closest to the GALEX centroid will take most of the flux- a reasonable assumption given that the {\bf SExtractor} object coordinates are based on flux moments. The advantage of this technique compared to redistributing flux based on $g$-band flux, or possibly even SEDs, is that incorrect matches that made it through the initial g-band selection will barely affect the GALEX UV magnitude since they will be more likely to lie further from the centroid in relative terms. If prior magnitude knowledge is used to make flux redistribution decisions this could introduce a significant bias into the derived colours.

\begin{figure*}
\begin{center}
\centerline{
\mbox{\includegraphics[width=3in]{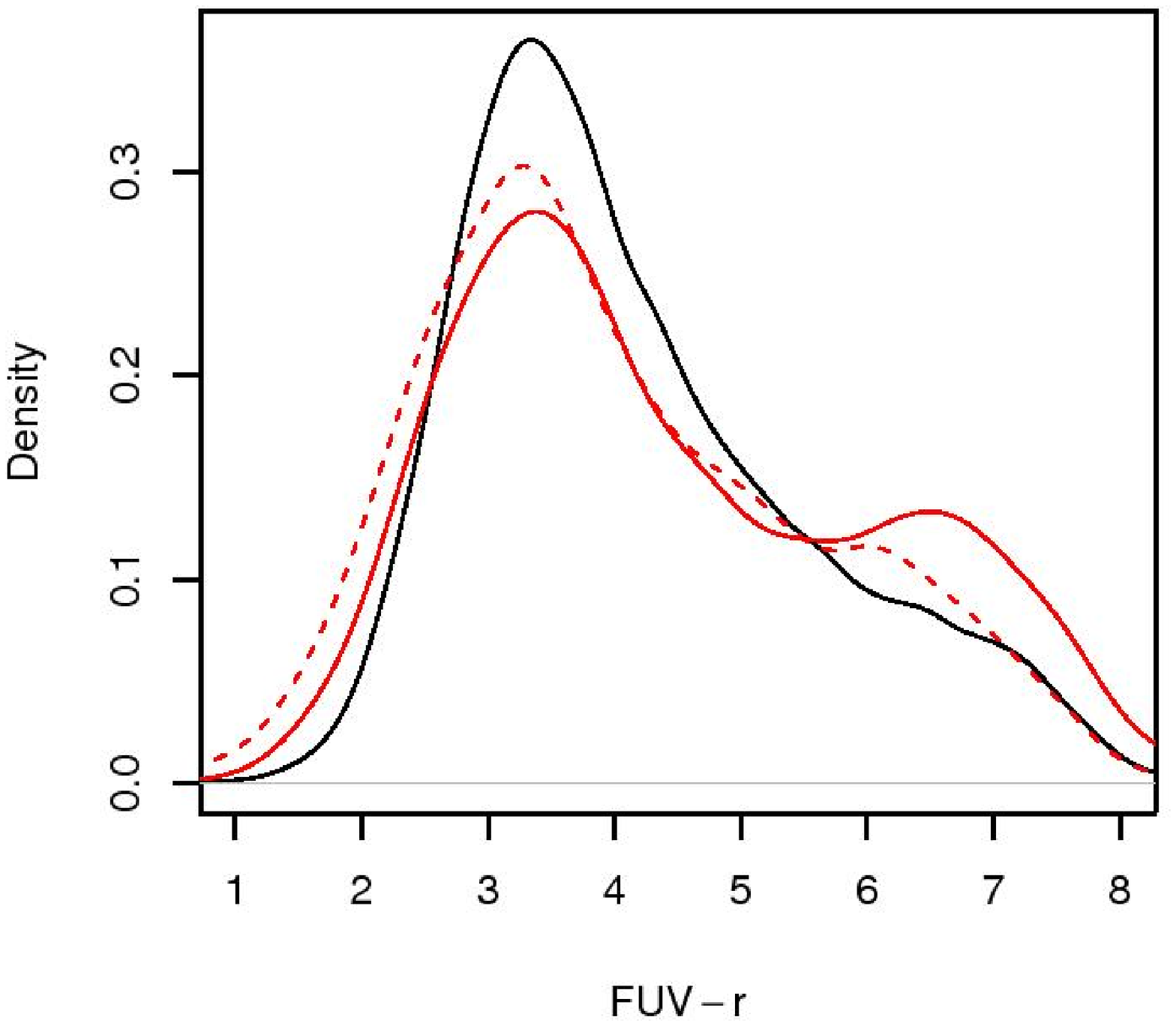}}
\mbox{\includegraphics[width=3in]{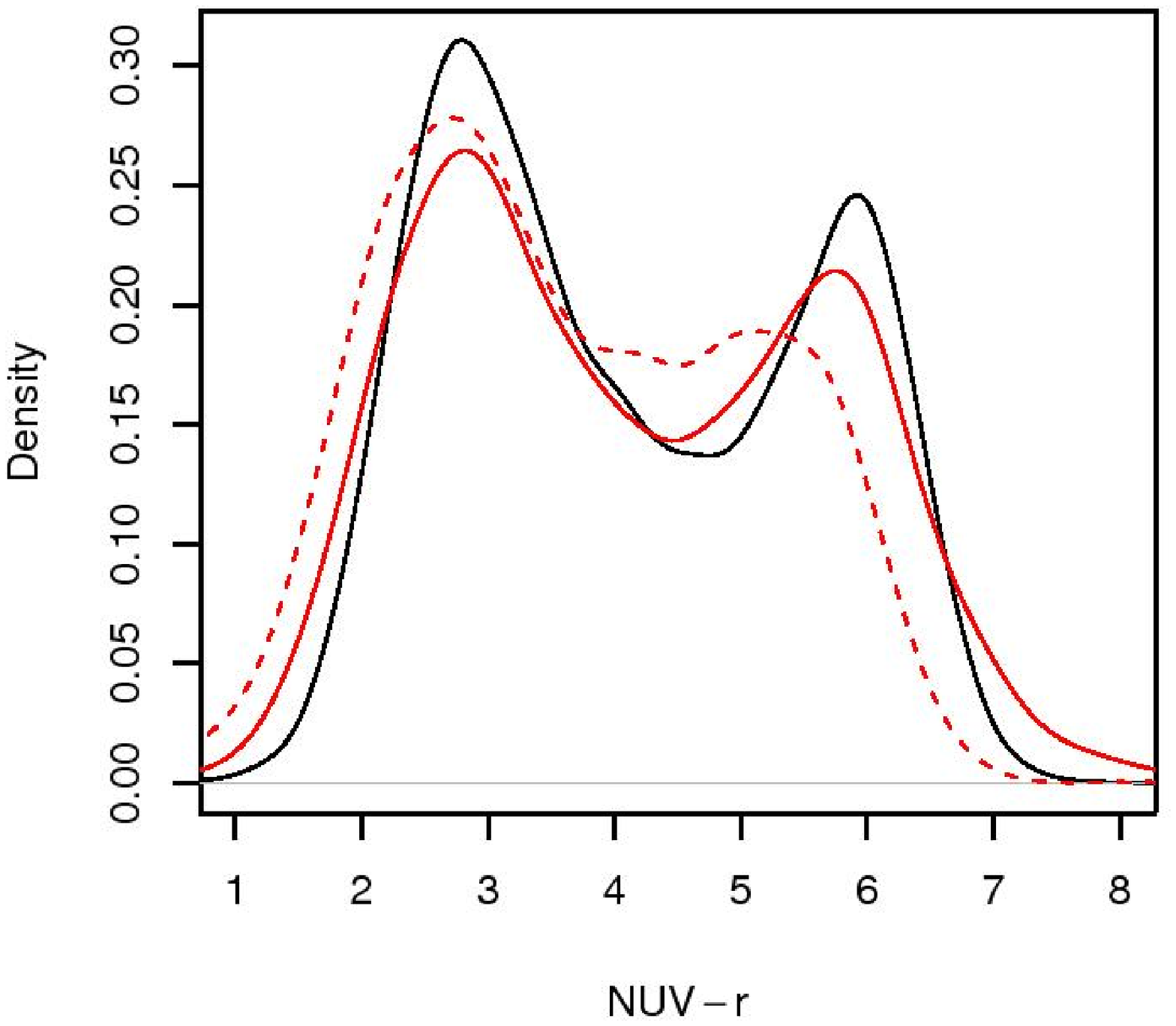}}
}
\caption{\small  Effect of the flux correction applied to ambiguous matching sources between GALEX and SDSS. The plots show how the correction affects the $FUV-r$ colour (left) and $NUV-r$ colour (right) for galaxies with redshifts in our sample. Black line shows the distribution for unambiguously matched SDSS-GALEX sources, the dashed red line shows the ambiguous sources before correction, and the solid red line after correction. It is particularly noticeable in the $NUV$ that the post corrected distribution is much closer to the unambiguous source matched distribution. The bimodality also strengthens significantly when the correction is applied, particularly evident for $NUV-r$ colours.}
\label{photoshift}
\end{center}
\end{figure*}

Using this simple technique of flux redistribution we find that the distribution of $NUV-r$ colours for ambiguous sources moves very close to those calculated for unambiguous sources (see Figure \ref{photoshift}). This is a very good indication that redistributing the flux in the manner described is both the right idea in theory, and has the desired empirical effect.

A plausible concern of the approach taken is that large objects might become broken up spuriously (despite the $g$ band selection designed to prevent this occurring). Considering objects that make the final sample and have an initial $NUV\le17$ (the brightest subset), $\sim40$\% of objects are associated with multiple SDSS objects that perhaps require some amount of flux. Of galaxies corrected, the median $NUV$ ($FUV$) magnitude shift is 0.31 (0.40), and in total this sample contains 98\% (99.4\%) of the total flux originally associated with it. For the full sample the corresponding figures are $NUV$ ($FUV$) magnitude shifts of 0.41 (0.47), and the whole sample contains 99.8\% (99.8\%) of the flux after flux redistribution. This would suggest that the flux for the brightest galaxies is not being reduced drastically compared to the full sample.

A side effect of the flux redistribution process is the creation of a very faint population of objects caused by erroneous distant objects taking a small amount of a large flux. In these cases, the primary object flux will barely change but a notable number of objects will gain a negligible flux. Since these magnitudes are far beyond the apparent magnitude limit used for producing the luminosity function, the effect is actually very minimal.

Overall 79.2\% of sources are unambiguous for our superset galaxy sample (SDSS objects flagged as galaxies), as defined by the matching mechanism outlined above. When applying the final survey selection cuts (discussed in detail later) $\sim93$\% of the sample is unambiguously matched (varying slightly on the GALEX band). It should be high-lighted that applying this flux correction does not have a dramatic effect on the final results presented here, the integrated fluxes remain almost the same. The main advantage of the approach advocated here would be for creating better quality SEDs for large samples of galaxies, and as such it has been applied to the multi-wavelength GAMA \citep{Driver2010b} survey which uses data from multiple PSF mis-matched sources.

Whilst the ultimate result of the flux correction method described appears to be better than just doing a simple nearest-neighbour match, there are a number of obvious caveats. Firstly, it assumes that the profile is circular. This will not always be the case, but for the majority of objects the 2D light profile convolved with the GALEX PSF will be approximately circular. This does mean that flux could be incorrectly redistributed to an off-axis object that is within the r$_{90}$ radius, when a simple match would not have done so. This is only likely to be a serious problem for the very largest objects r$_{90}>10$'', and these objects do not suffer unusually large amounts of flux loss in the redistribution process, suggesting the effect is not likely to be dominant. Another limitation is that this method does not account for colour in an explicit manner: galaxies which are very blue in $g-r$ are likely to deserve more missing UV flux than galaxies which are very red in $g-r$. There are some indirect limits imposed by the requirement that the additional objects be within 2.5 mag of the primary match in the g-band, but otherwise there is no colour restriction in the implementation described. The use of colour (and by extension full SEDs) was considered, but there was a desire to keep the algorithm as simple as possible. Figure \ref{photoshift} suggests that the red population is the one most overtly affected by flux loss in GALEX, and in fact more flux should be shifted from the blue to red populations. Despite the limitations discussed, flux shifting appears to produce a better fidelity of data than simply matching via nearest-neighbour.

\subsection{Completeness}

\begin{figure}
\centerline{
\includegraphics[width=3in]{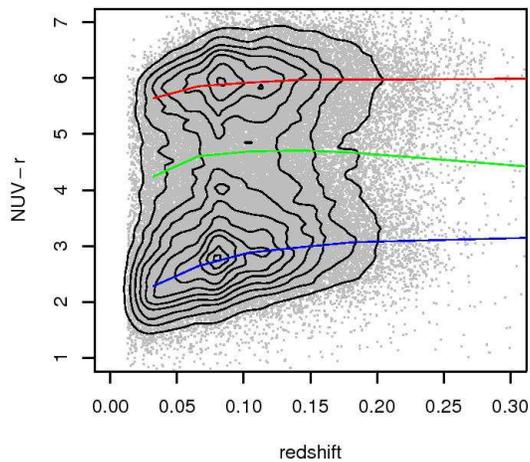}
}
\caption{\small  $NUV - r$ as a function of redshift. The blue line represents the peak in the bimodality for blue (late-type) galaxies, the red line represents the peak in the bimodality for red (early-type) galaxies and the green line represents the minima between the two peaks. The linear trend seen for blue (late-type) galaxies over this redshift regime is just a selection effect caused by sampling lower luminosity (more numerous) objects at lower redshifts for a fixed blue colour.}
\label{splitcontour}
\end{figure}

To calculate the completeness function required to generate galaxy weights the SDSS star galaxy classification was adopted to identify potential targets that could have been selected by virtue of their $r$-band luminosity ($r_{\rm petro}<17.77$mag). GALEX target completeness was identified as a function of $r_{\rm petro}$ in the early and late type populations separately. To split the populations the $NUV-r$ colour bimodality was studied as a function of redshift, and the minimum in the bimodality was used to select late-type star forming populations and redder colours to select eary-type quiescent populations (see Figure \ref{splitcontour}). The dichotomy is significant in terms of completeness, with late-type galaxies possessing a higher completeness by virtue of their easily identifiable emission features aiding redshift measurements. Over the vast majority of galaxies the completeness is quite flat as both a function of $NUV$ or $r$, where the average completeness for early-type galaxies is $\sim78$\%, for late-type galaxies it is $\sim84$\% and globally it is $\sim82$\%. The inverse of the completeness as a function of $r_{\rm petro}$ and galaxy type is used to generate weights for each galaxy. Assuming a constant a magnitude dependent weight does not have a large impact on any of the results presented here.

\subsection{Photometric Corrections}

The GALEX photometry acquired from MIS required Galactic extinction corrections and K-corrections. For the former $E(B-V)$ values calculated using the Schlegel (1998) dust maps. The reddening factor for GALEX bands is $A_{FUV}/E(B-V)=8.24$ mag and $A_{NUV}/E(B-V)=8.2$ mag \citep[$NUV$ has a slight spectral dependence, but we use the constant values used in][]{Wyder2007}, which due to our low extinction plate requirement means the mean correction is less than 0.34 mag in both bands.

\begin{figure*}
\centerline{
\mbox{\includegraphics[width=3in]{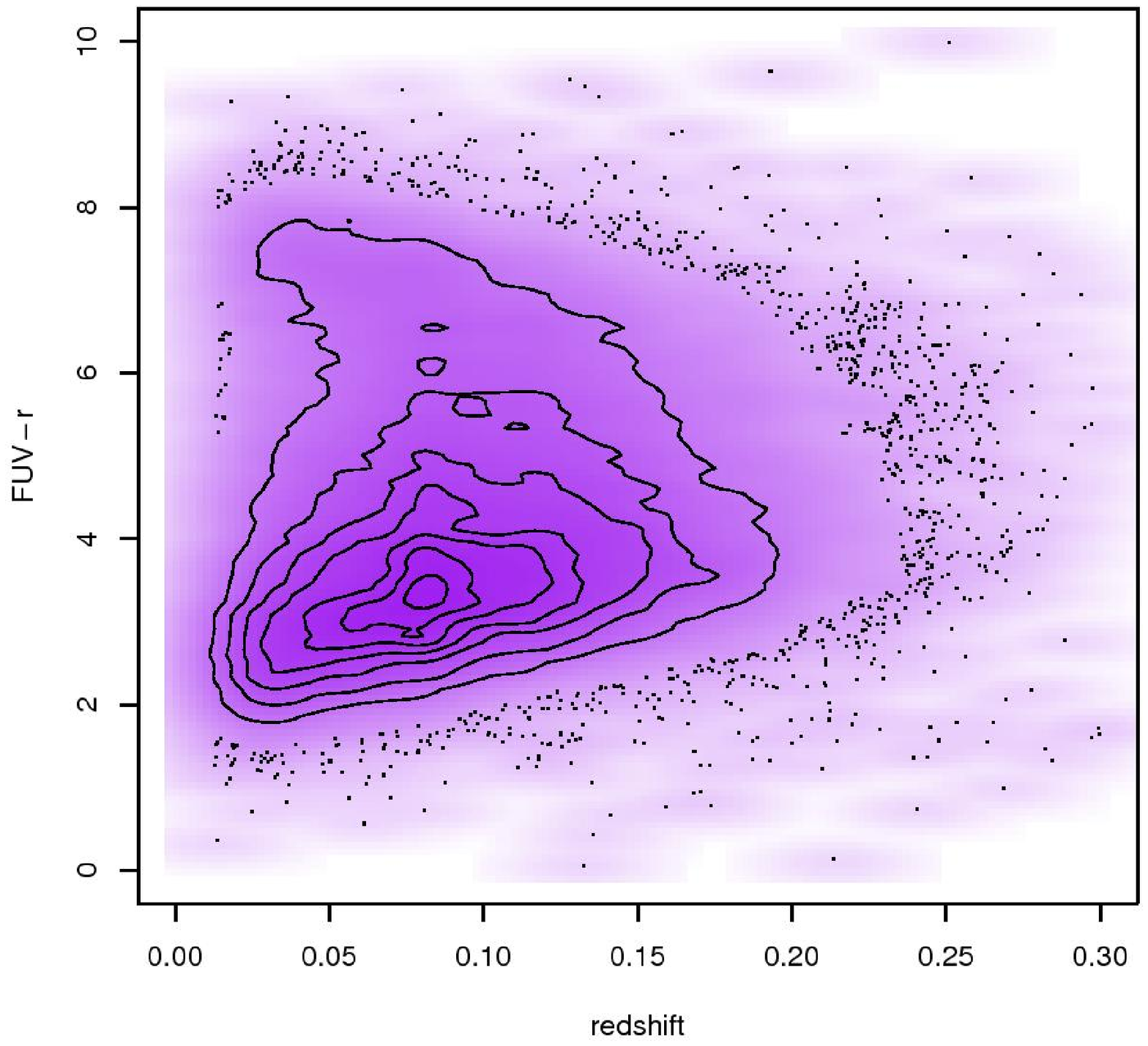}}
\mbox{\includegraphics[width=3in]{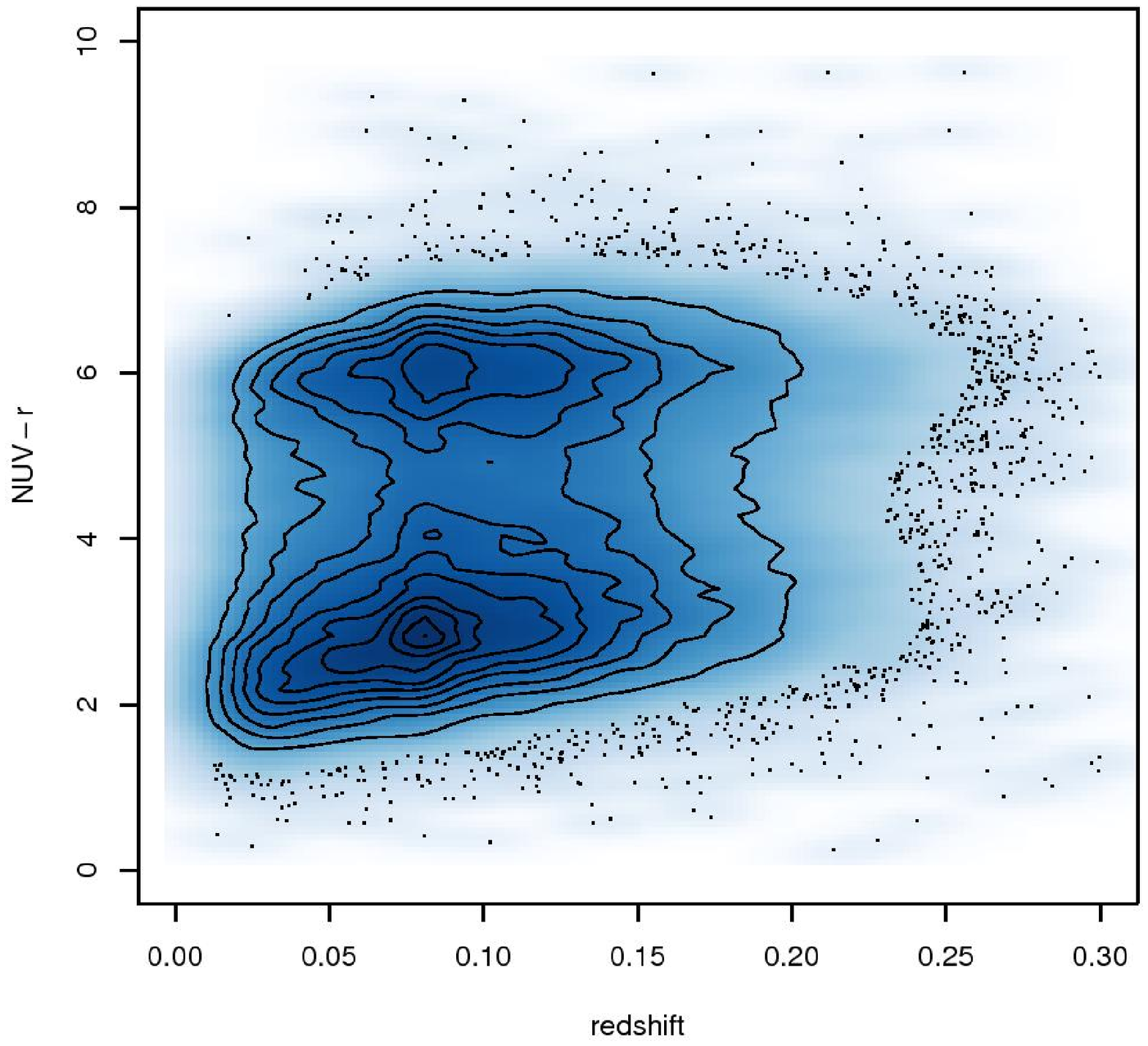}}
}
\centerline{
\mbox{\includegraphics[width=3in]{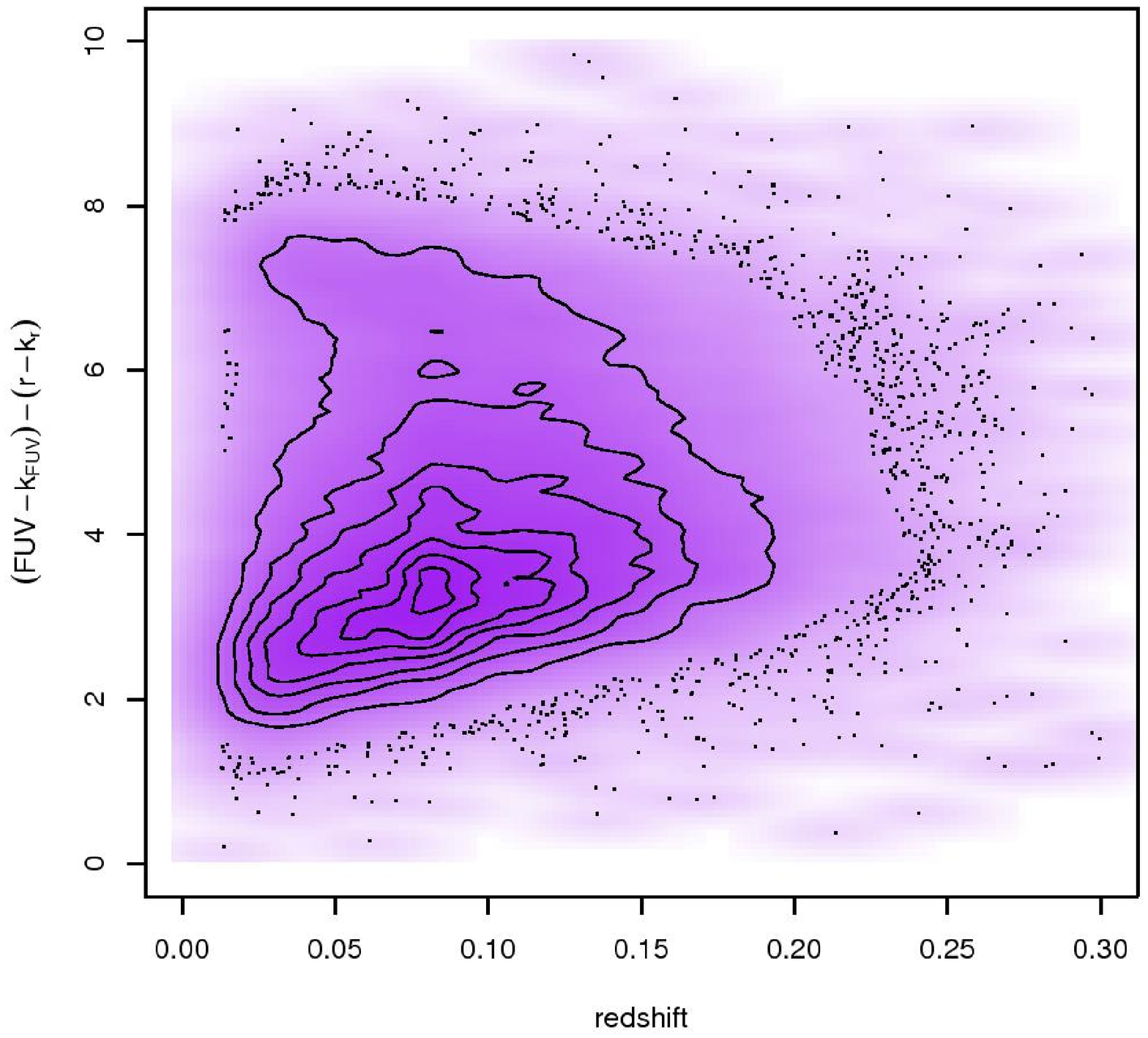}}
\mbox{\includegraphics[width=3in]{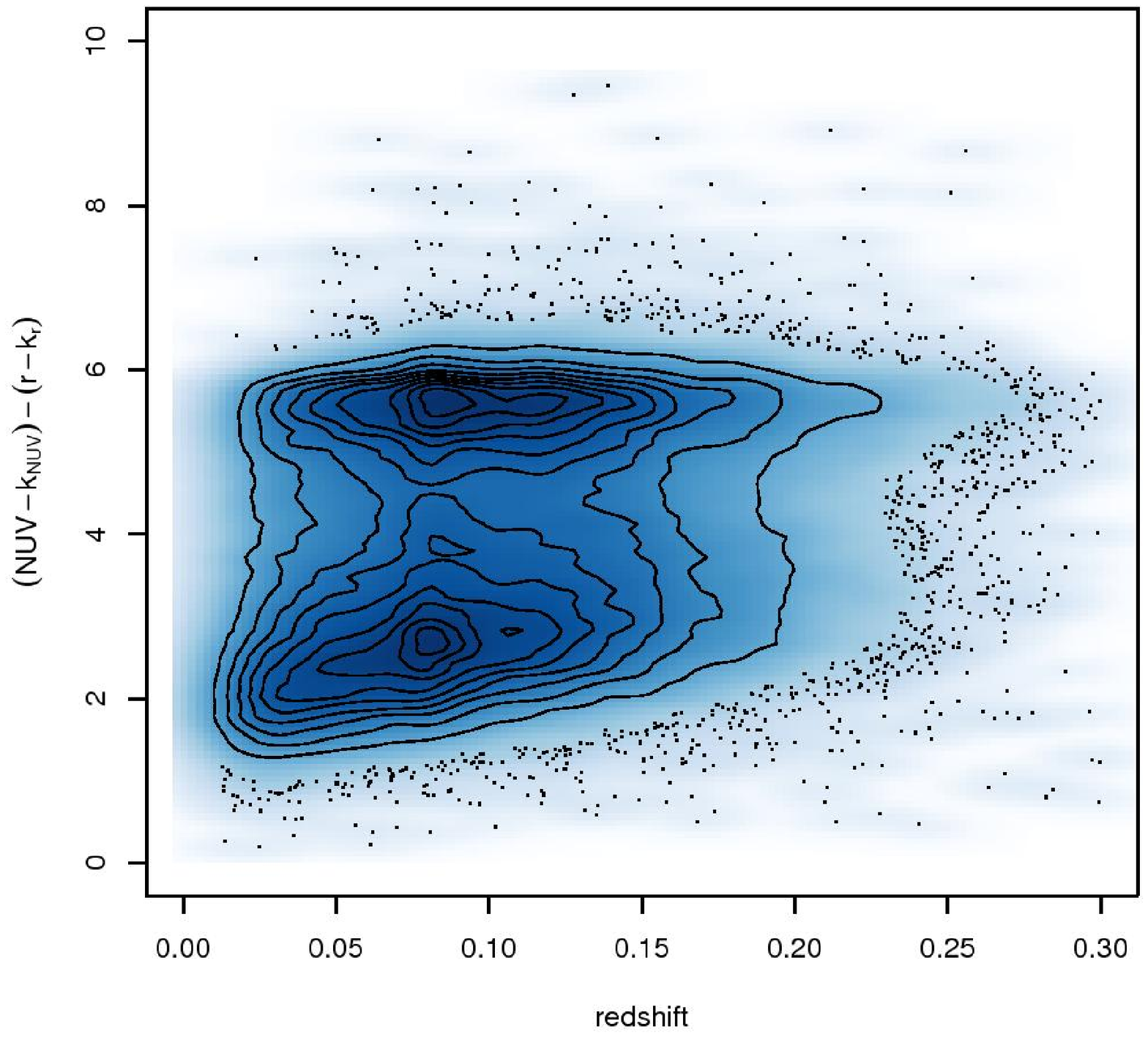}}
}
\caption{\small  Comparison of $FUV-r$ and $NUV-r$ colours before (upper panels) and after (lower panels) k-corrections. The 1,000 points in the least dense part of the parameter space are plotted, for the rest the intensity of colour is a guide to the local density of points. The significant observation is how the bimodality strengthens and the redshift dependence disappears post correction.}
\label{kcorrect}
\end{figure*}

The k-corrections applied were created using {\bf kcorrect-V4.2} \citep{Blanton2007a}, which fits the dust corrected photometric SED for each galaxy ($FUV_{\rm kron}$, $NUV_{\rm kron}$, $u_{\rm petro}$, $g_{\rm petro}$, $r_{\rm petro}$, $i_{\rm petro}$, $z_{\rm petro}$ were the bands used in this work, and should be presumed throughout) to a library of templates and calculates the change in flux measurement due to the the filters shifting and stretching compared to observing the same galaxy at $z=0$. Figure \ref{kcorrect} shows the effect of applying the k-correction to the $FUV$ and $NUV$ bands as a function of redshift. The most noticeable effect is that the early-type/ late-type bimodality most evident in $NUV-r$ becomes more distinct. The $NUV-r$ colour for redder (early-type) galaxies is much more distinct, and appears to remain very flat as a function of redshift. The bluer (late-type) galaxies in the $NUV-r$ plot appear to show strong evidence for evolution in colour over the baseline of 3.4 Gyr (out to $z=0.3$) shown in this plot, but this effect is actually a selection effect caused by the fixed $r$ and $NUV$ limits. Crudely speaking an object with a $NUV-r\sim 2$ with $r=17.77$ (i.e.\ at the SDSS limit) will have M$_{\rm NUV}$=-13.21 at $z=0.013$, whereas at $z=0.3$ the same magnitudes correspond to M$_{\rm NUV}$=-20.41. Assuming a typical distribution of absolute magnitudes, we would expect fewer galaxies to be present at $z=0.3$ for the same $NUV-r$ colour. We do not attempt to correct for evolution in the sample, the corrections being small over the $\sim$1 Gyr baseline investigated here.

\subsection{Sample Limits}

\begin{figure*}
\centerline{
\mbox{\includegraphics[width=3in]{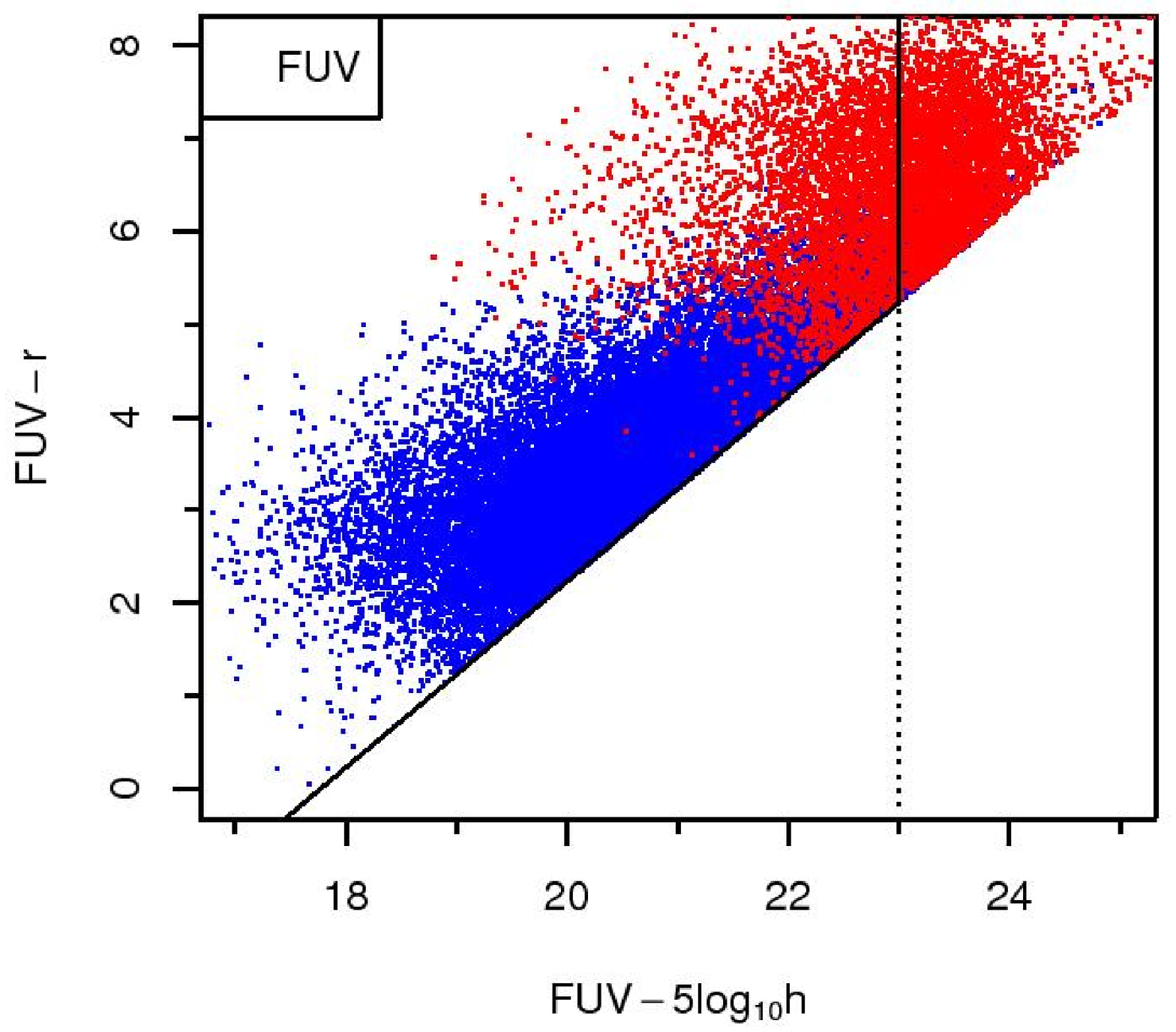}}
\mbox{\includegraphics[width=3in]{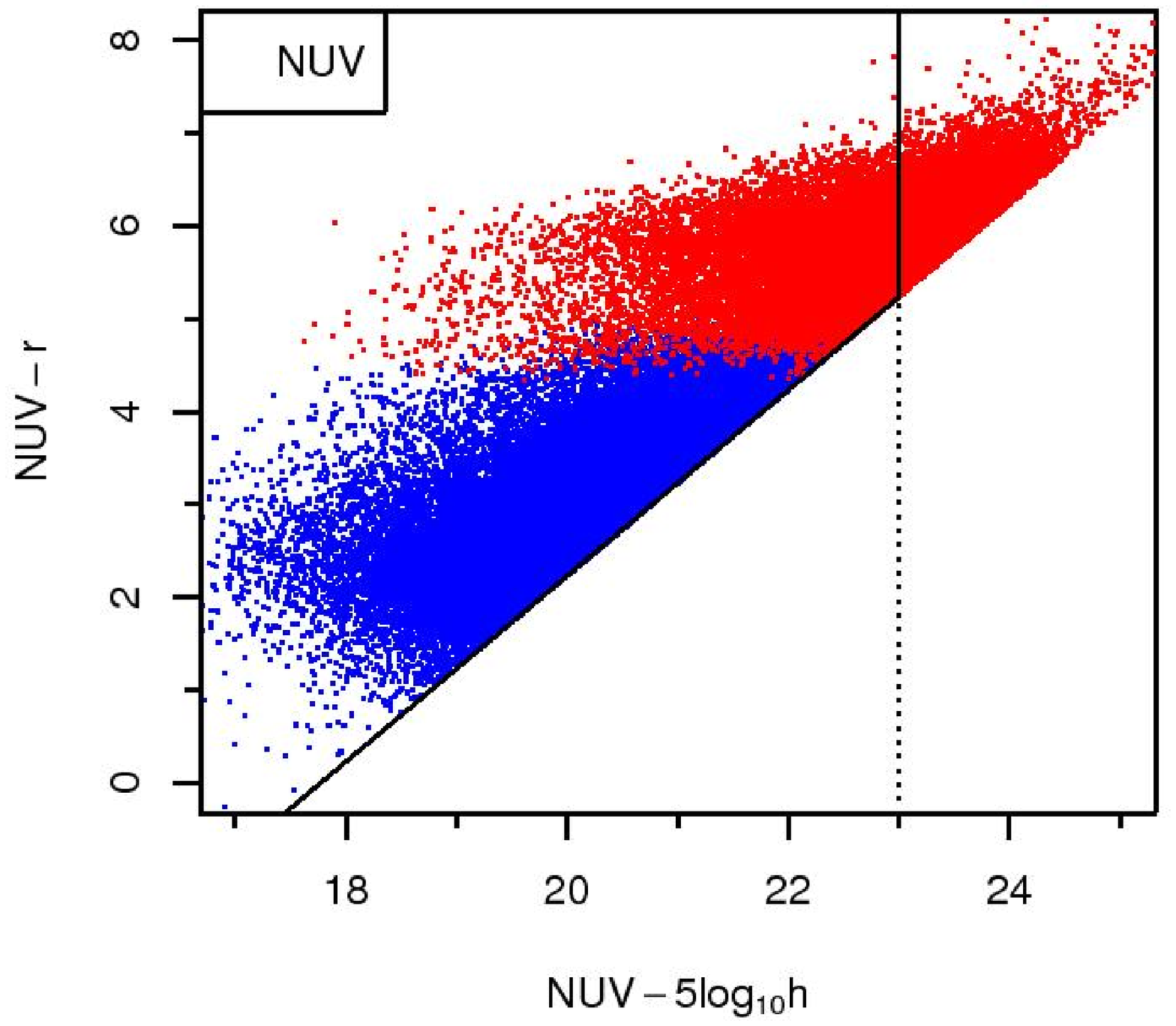}}
}
\caption{\small  Comparison of $FUV$ and $NUV$ survey depths. The effective $r_{petro}$ limit is 17.77 mag throughout, and the assumed maximum GALEX limits are 23.0 for $FUV$ and 23.0 for $NUV$ (these are the dotted lines drawn). The floating survey limit in each GALEX band is that implied by the solid black line, where the GALEX colour ($FUV-r$ or $NUV-r$) and the GALEX apparent magnitude define the survey limit.}
\label{depth}
\end{figure*}

Choosing a sensible depth for the survey, and mapping this onto an observable absolute magnitude limit, is important. The process is made more complicated by virtue of the redshift catalogue survey (SDSS) is limited by $r_{\rm petro}<17.77$ mag, whilst GALEX is photometrically limited in the $FUV$ and $NUV$ bands. The colour magnitude diagrams for $FUV-r$ versus $FUV$ and $NUV-r$ versus $FUV$ are shown in Figure \ref{depth}. The assumed limits are 23.0 mag for both $FUV$ and $NUV$, this is brighter than the 23.8 mag and 23.6 mag depths given in \citet{Xu2005} for $FUV$ and $NUV$ so should be a conservative estimate.

Due to the limited depth of the SDSS redshift survey (median redshift is $z \sim 0.1$), and the desire to sample a minimal amount of evolution so as to obtain flux densities for the most local Universe, we used a redshift sample spanning $0.013 \le z \le 0.1$, where the low redshift cut is used to avoid nearby galaxies that require significant local flow corrections, and the higher redshift limit is the same as used in \citet{Wyder2005}. This makes comparison more straightforward, and keeps the measured UV flux within a 1 Gyr baseline.
 
\section{Results}

The Luminosity Distributions (LDs) for GALEX were generated using a standard $1/V_{\rm max}$ method \citep[in a manner similar to][]{Wyder2007} for 10 subsets of the sample split by $NUV-r$ colour. The subsets were determined by calculating the even quantiles of the data. To minimise colour bias within each subset of the data we calculated the $10^{th}$ percentile of each subset and use this as our effective magnitude limit for $NUV$/$FUV$, leaving the $r$-band limit fixed at $r_{\rm petro}\le17.77$ mag. Attempts were made to split the sample up into either simply late/early type and late/central/early type, but the colour bias was such that the resulting LFs were heavily distorted. Using the colour binning specified was a good compromise between data size for each subset and quality of the resulting LD (i.e.\ by visual inspection it appeared physical). Such an approach also means we keep the maximum possible number of galaxies when calculating the total integrated luminosity density of the local Universe.

The consequence of dividing the sample like this is that we are not left with a simple blue (late-type) and red (early-type) LDs. To determine appropriate luminosity densities for the full sample, and the early/late-type populations, stacking of these distributions must be done. Since we are interested in the integrated light the best approach is to fit a parametric function to the data subsets and to integrate under this function beyond the depth we can observe to. Since we already know the fraction of early/late-type galaxies belonging to each colour subset, we can share the integrated light between the two populations to calculate the luminosity density for the early-late-type and composite populations separately.

\subsection{Luminosity Functions}

To these empirical LDs we fitted the analytic Schechter Luminosity Function (LF), a 3 parameter function that is widely used in the literature, and allows for simple comparisons between models to be made \citep{sche76}:

\begin{equation}
\frac{dn}{dL}=\phi(L)=\frac{\phi_{*}}{L_{*}}e^{-L/L_{*}}\left(\frac{L}{L_{*}}\right)^{\alpha}.
\end{equation}

A more complicated form of this function can be defined in terms of magnitudes ($M^{*}$ and $M$) in place of luminosities ($L^{*}$ and $L$). The 3 parameters that we fit are $M^{*}$, which for most distributions represents the magnitude of the object that provides the most overall flux; $\phi^{*}$, approximately the comoving space density of galaxies with the magnitude $M^{*}$ and acts as a normalisation constant; and $\alpha$, a power law term that describes the faint end slope of the LF. The combination of all 3 parameters allow us to calculate the energy output per unit volume. Even though the faint end will truncate eventually, the flux output is usually dominated by galaxies near $M^{*}$, so the important consideration is that we sample beyond $M^{*}$ to ensure this is well measured. In practice ~1 mag is the minimum beyond $M^{*}$ that should be sampled. To fit the Schechter function we find the minimum $\chi^{2}$ fit for each distribution, where the errors are calculated using the covariance matrix produced during the fit.

\begin{figure*}
\centerline{
\mbox{\includegraphics[width=3in]{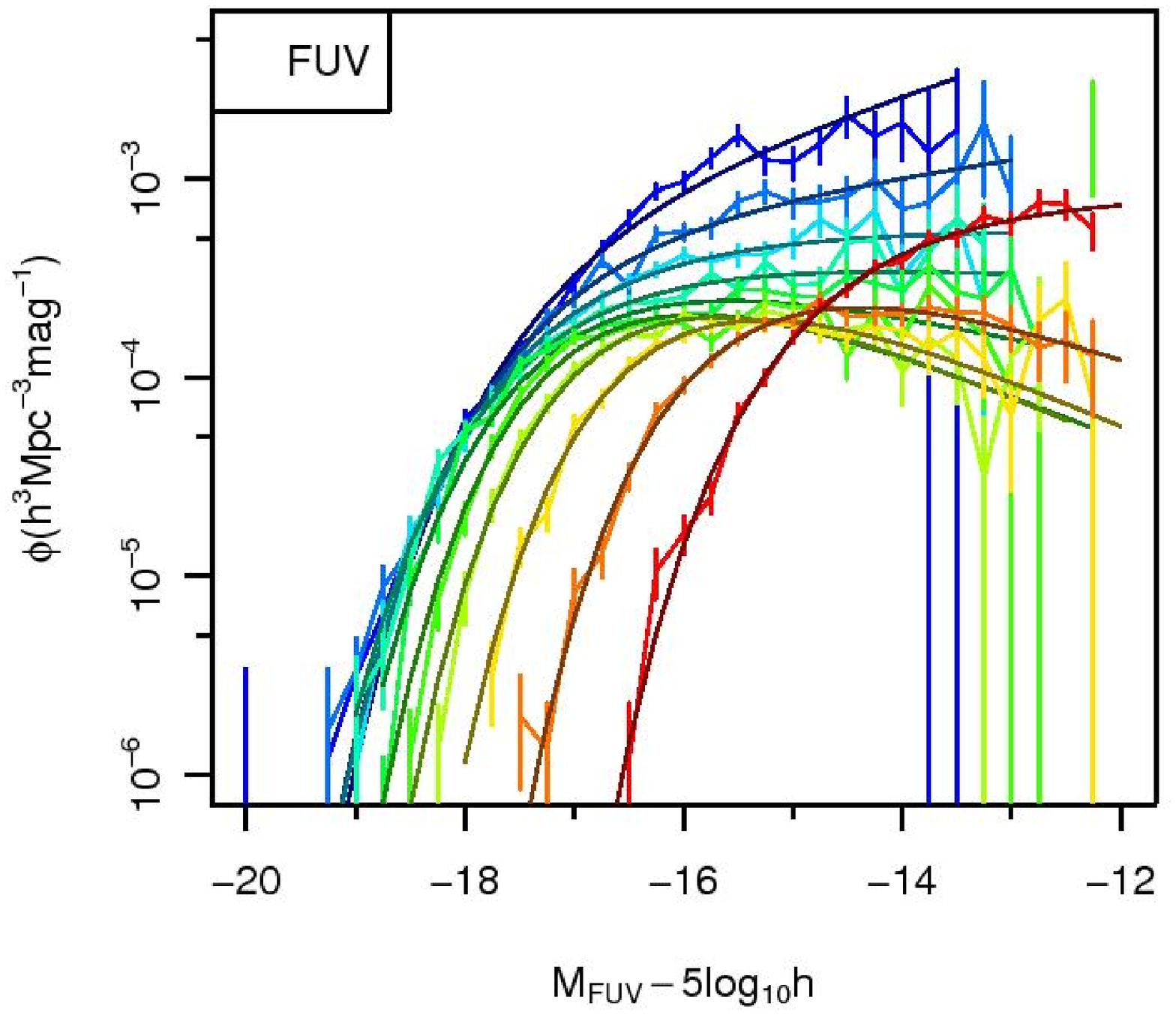}}
\mbox{\includegraphics[width=3in]{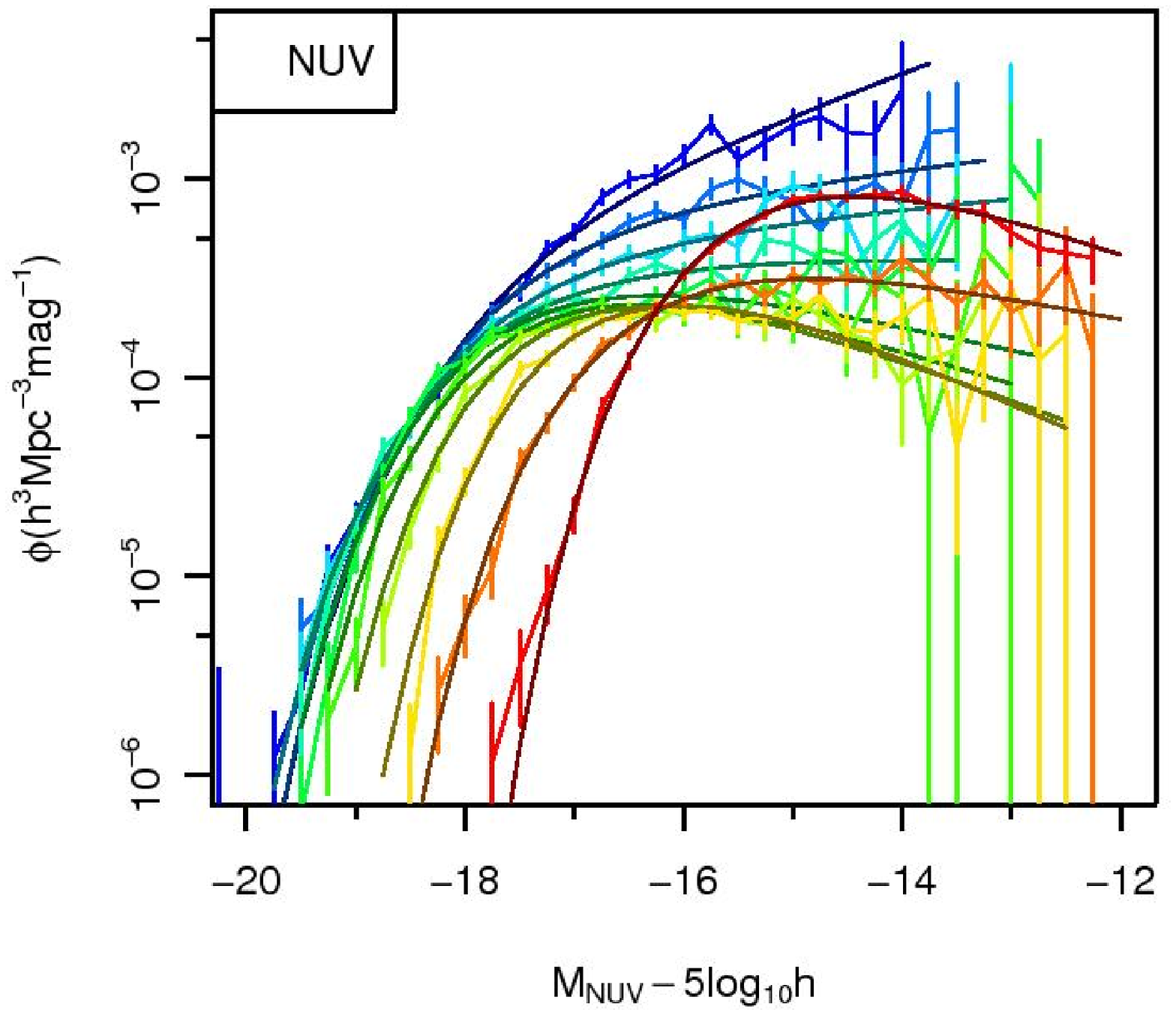}}
}
\caption{\small  Comparison of $FUV$ and $NUV$ luminosity functions for early-type, late-type and composite populations of galaxies. The blue--green--red colour scaling maps maps onto the bluest (lowest) $FUV-r$ ($NUV-r$) to the reddest (largest) $FUV-r$ ($NUV-r$).}
\label{LFs}
\end{figure*}

The empirical LDs created via $1/V_{\rm max}$ and the best fit Schechter LFs can be seen in Figure \ref{LFs} for the $0.013 \le z \le 0.1$ sample. The full table of Schechter parameters and derived values is presented in Tables \ref{FUVparams} and \ref{NUVparams}. The median $\chi^{2}/\nu$ for each fit is 1.07/1.02 for $NUV$ and $FUV$ data respectively, demonstrating that the subset divisions used are appropriate since the data are neither under or over-fitted. Figure \ref{errcon} shows the 1, 2 and 3$\sigma$ error contours for the colour subsets used in this work. A dramatic feature of the LFs is that they maintain a relatively constant $M^{*}$ for the first five colour bins for both $FUV$ and $NUV$ ($M^{*}_{FUV} \sim -17.2$ mag and $M^{*}_{NUV} \sim -17.7$ mag) but with a rapidly increasing (-1.5 -- -0.7) $\alpha$. The value of $\alpha$ reaches a maximum value for the $7^{th} / 8^{th}$ colour subsets combined with a rapidly faintening $M^{*}$, before $\alpha$ appreciably steepens again for the final few reddest bins.

\begin{figure}
\centerline{
\includegraphics[width=3in]{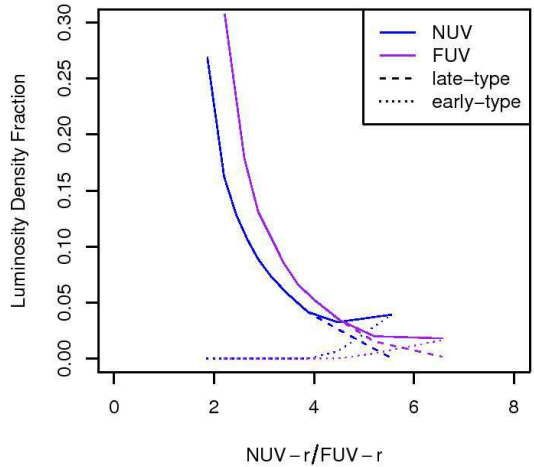}
}
\caption{\small  Fraction of total luminosity density contained for each colour subset. Dashed and dotted lines show the late/early-type contribution separately, where, as should be expected, the late-type population dominates the bluer colours. Overall, the vast fraction of integrated UV light ($FUV$ and $NUV$) comes from the bluer subsets/ late-type populations.}
\label{lumdenfrac}
\end{figure}

The transitions in $M^{*}$ and $\alpha$ are statistically significant--- the major parameter shifts actually occur orthogonally to the main direction of $M^{*}$ and $\alpha$ degeneracy (the vector of ellipse elongation seen in Figure \ref{errcon}). Together with the shifts in $\phi^{*}$, these fits mean the integrated flux varies hugely as a function of $FUV-r$ of $NUV-r$ colour. The immediate observation is that the bluest colour bins contain the vast majority of the integrated flux, this is particularly clear in Figure \ref{lumdenfrac} which shows the fractional contribution of each colour subset to the total integrated light. The early/late-type contributions are also shown, where 97.9\% (95.5\%) of $FUV$ ($NUV$) light is found in late-type galaxies in the local Universe.

\begin{figure*}
\centerline{
\includegraphics[width=3in]{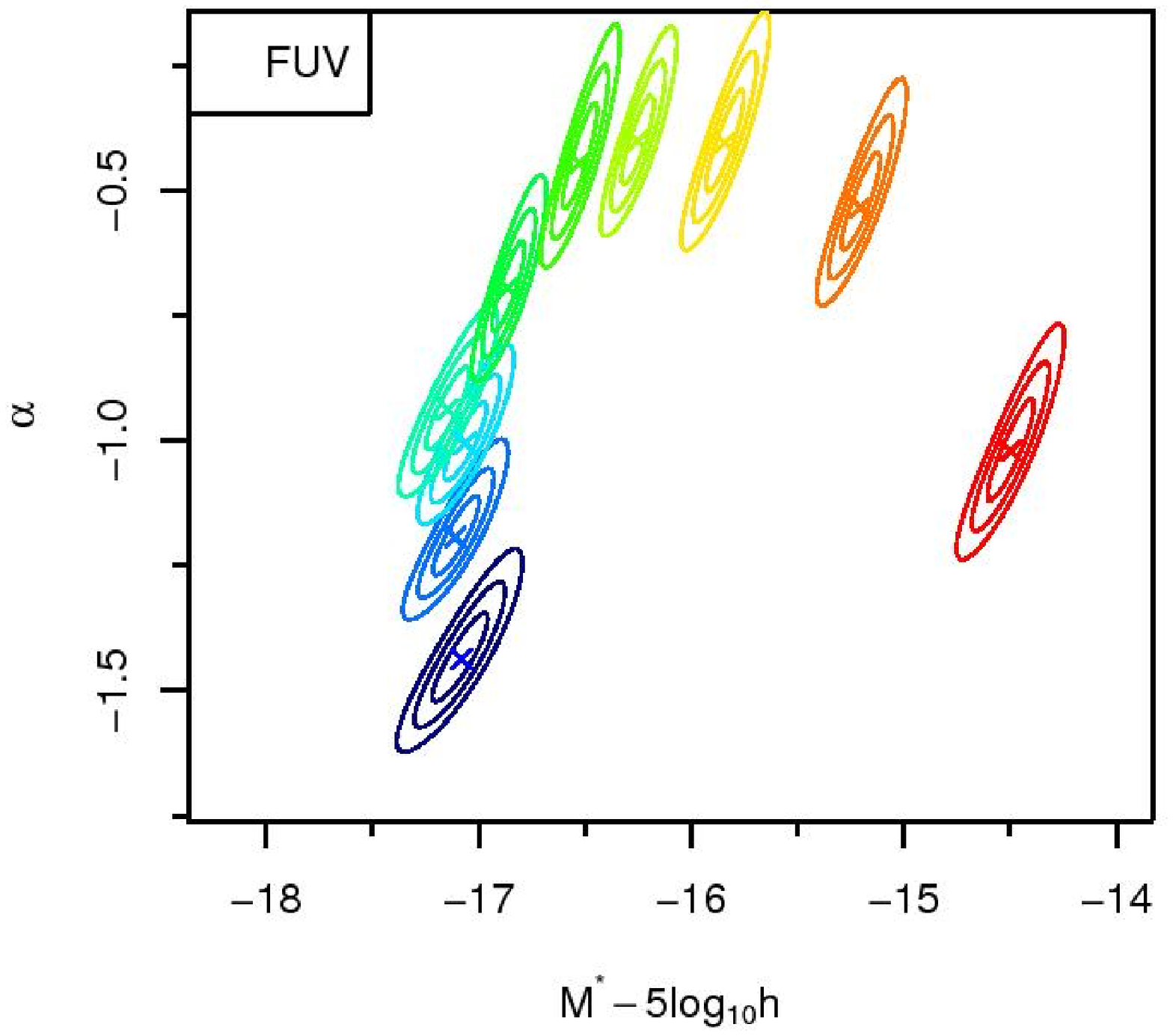}
\includegraphics[width=3in]{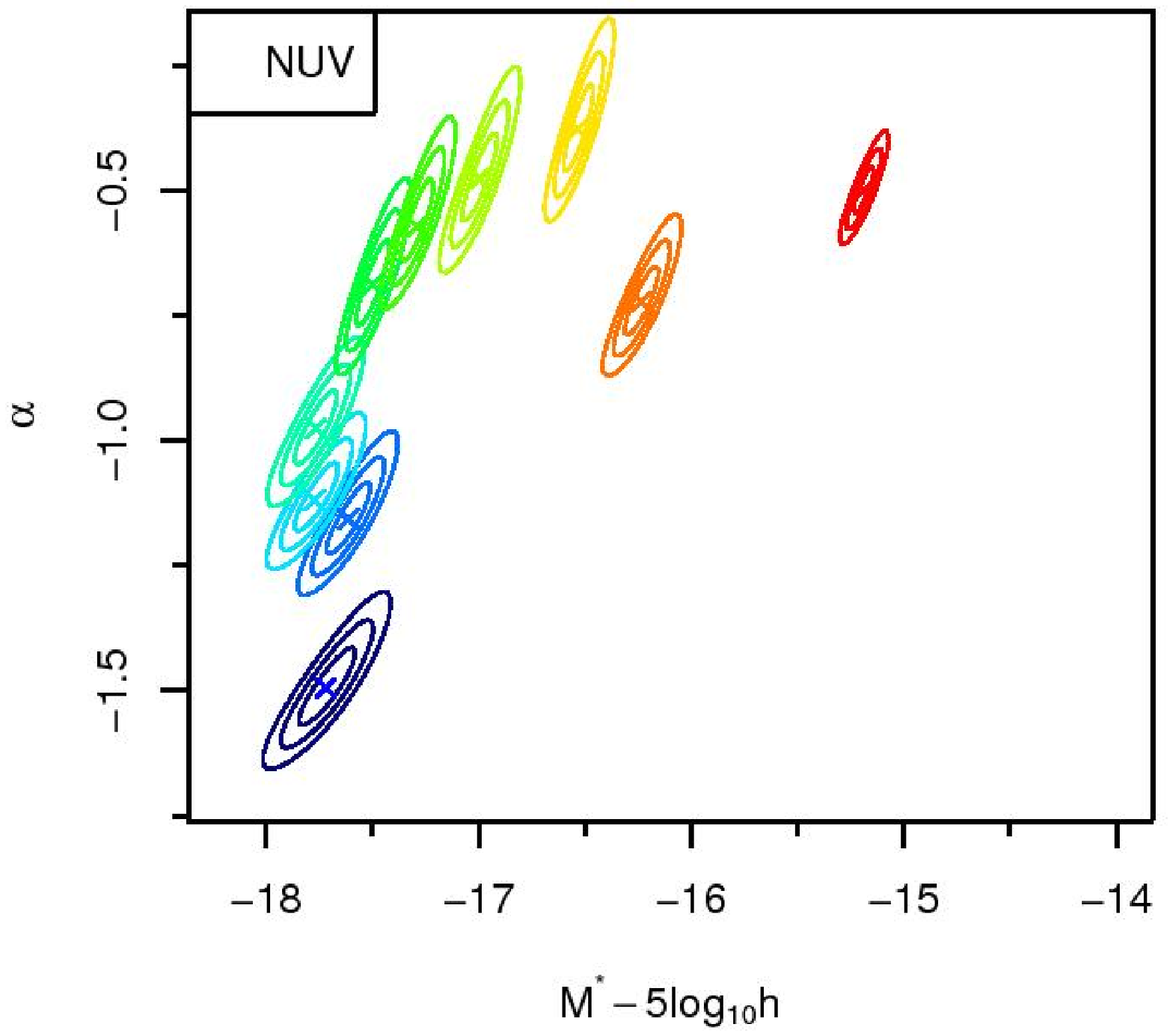}
}
\caption{\small  Comparison of $FUV$ (left) and $NUV$ (right) luminosity functions for the 10 different colour subsets given in Tables \ref{FUVparams} and \ref{NUVparams}. The blue--green--red colour scaling maps maps onto the bluest (lowest) $FUV-r$ ($NUV-r$) to the reddest (largest) $FUV-r$ ($NUV-r$). }
\label{errcon}
\end{figure*}

The $FUV$ and $NUV$ LFs presented here have similarities to those in \citet{Wyder2007}. Figure 19 in that paper can be compared almost directly with the $NUV$ LFs in Figure \ref{LFs}. Both demonstrate a significant faintening of $M^{*}$ and a flattening of $\alpha$ as the subsets transition from blue to red populations. \citet{Wyder2007} use regular colour subsets to divide the populations, whereas here the number of galaxies used to construct each LF is designed to be as similar as possible, as such the most extreme red and blue populations investigated in that work are not present here.

\subsection{Composite Luminosity Function}

To allow for easy comparisons to further work the full composite galaxy densities and errors have been calculated using 0.25 mag bins spanning -20.25 to -14.00 for both $FUV$ and $NUV$ data, using the 10 colour split populations shown in Figure \ref{LFs}. These are presented in the Appendix, and include summed versions of the early-type and late-type population samples for each magnitude bin. The preference is that these empirical values should be used in any comparison, but for completeness the best composite $FUV$ Schechter function has $\phi^*=0.0165 \pm 0.0007 h^{3}$Mpc$^{-3}$mag$^{-1}$, $M^*=-17.031 \pm 0.030$ and $\alpha=-1.167 \pm 0.025$ and the best composite $NUV$ Schechter function has $\phi^*=0.0169 \pm 0.0006 h^{3}$Mpc$^{-3}$mag$^{-1}$, $M^*=-17.641 \pm 0.025$ and $\alpha=-1.200 \pm 0.019$.

\subsection{Luminosity Densities}

The most interesting derived components, given a well behaved LF that can be parameterised accurately with a Schechter fit to the distribution, is the total integrated light output of galaxies per unit volume and the implied star formation density (SFR) of the Universe. Following the approach taken in many studies \citep[e.g.][]{Madau1996,Baldry2003,Driver2008,Hill2010}, we can integrate the implied Schechter function out to infinity using the $\Gamma$ function:

\begin{equation}
f_{\nu}=h (10^{-(M^{*}-2.5 \log_{10}(\phi^{*}\Gamma(2+\alpha)))/2.5}) 4.345 \times 10^{20}.
\end{equation}

This will give the Universal flux density in units of ergs s$^{-1}$ Hz$^{-1}$ Mpc$^{-3}$, and because of the cancelation between absolute magnitude (which depends on distance squared) and density (distance cubed) is only dependent on a single factor of $h$, assuming otherwise standard cosmology. As discussed above, the influence of the faint-end is relatively minor for typical $\alpha$ values. If the empirical LF extends 4 magnitudes faintwards of $M^{*}$ then integrating out to infinitely faint fluxes rather than using the limit implied by the data only increases the flux density by 2.5\% for $\alpha=-1$ and 13.8\% for $\alpha=-1.3$. At worst, this effect is of similar importance as the cosmic variance in the sample. The role of cosmic variance (CV) for different surveys, and in particular different survey geometries, is discussed in some detail in Driver \& Robotham et al. (2010). Taking our redshift range of $0.013 \le z \le 0.1$ and the area coverage of 833.13$^{\circ}$ we have a total survey volume of 2.112 Gpc$^{3}$h$^{-3}$. Assuming the worst case scenario of our survey having equal dimensions the formula presented in Driver \& Robotham (2010) predicts a CV$\sim 12.6\%$, and no lower than 12.0\% assuming geometry similar to that actual used. All densities, and values derived from densities, can be considered to possess this extra form of error.

The derived parameters, $f_{\nu}$ and the SFR (determined using \citep{Kennicutt1998} $SFR_{FUV}$($M_{\odot}$yr$^{-1}$)$=1.4\times10^{-28}L_{\nu}$(ergs s$^{-1}$ Hz$^{-1}$) ), are largely consistent with those found by \citep{Wyder2005} and \citep{Budavari2005}, the appropriate conversions to the units used in this paper are in Tables \ref{FUVparams} and \ref{NUVparams}. The flux density in the $FUV$ agrees within implied errors for both the low redshift \citep{Wyder2005} sample and the higher redshift \citep{Budavari2005}. There is more of a discrepancy for the $NUV$ where we find $f_{NUV}=9.706 \pm 0.541 \times 10^{25}$ ($h$ ergs s$^{-1}$ Hz$^{-1}$ Mpc$^{-3}$) and \citet{Wyder2005} find $f_{NUV}=7.52 \pm 2.16 \times 10^{25}$ ($h$ ergs s$^{-1}$ Hz$^{-1}$ Mpc$^{-3}$). This is only slightly larger than the formal errors calculated from integrating the LFs. A possible cause for this difference is that the \citet{Wyder2005} redshift sample was constructed from matching the overlap between the GALEX MIS and the Two Degree Galaxy Redshift Survey \citep[2dFGRS;][]{Colless2001}, a completely separate volume to the one considered here and in \citet{Budavari2005}. This means differences arising due to CV effects are expected to be more significant since there is no common volume probed in the two studies, an effect that is discussed in detail later. The $NUV$ flux density estimate in \citet{Wyder2007} is much closer to that presented here: $f_{NUV}=8.83 \pm 0.61 \times 10^{25}$ ($h$ ergs s$^{-1}$ Hz$^{-1}$ Mpc$^{-3}$). This later paper uses SDSS data also, so a convergence in results should be expected since a large fraction of the data will be the same between the two analyses.

\begin{table*}\tiny
\begin{center}
\begin{tabular}{lccccccc}
  \hline
 $FUV-r$ & $\phi^*$ & $M^*-5 \log_{10}h$ & $\alpha$ & $f_{\nu} \times 10^{25}$ & $\chi^{2} / \nu$ \\ 
 & ($h^{3}$Mpc$^{-3}$mag$^{-1}$) & (mag) &  & ($h$ ergs s$^{-1}$ Hz$^{-1}$ Mpc$^{-3}$) & \\ 
  \hline
2.215&$0.00345\pm0.00046$&$-17.082\pm0.086$&$-1.436\pm0.058$&$1.607\pm0.288$&1.835\\
2.607&$0.00266\pm0.00026$&$-17.111\pm0.072$&$-1.193\pm0.052$&$0.936\pm0.114$&1.604\\
2.881&$0.00236\pm0.00018$&$-17.064\pm0.066$&$-1.005\pm0.052$&$0.687\pm0.073$&0.506\\
3.134&$0.00187\pm0.00014$&$-17.146\pm0.068$&$-0.938\pm0.055$&$0.569\pm0.059$&0.963\\
3.390&$0.00207\pm0.00011$&$-16.862\pm0.051$&$-0.695\pm0.060$&$0.448\pm0.031$&2.424\\
3.683&$0.00219\pm0.00009$&$-16.526\pm0.054$&$-0.446\pm0.070$&$0.345\pm0.023$&2.207\\
4.040&$0.00217\pm0.00008$&$-16.253\pm0.052$&$-0.402\pm0.061$&$0.268\pm0.017$&1.080\\
4.532&$0.00210\pm0.00009$&$-15.843\pm0.061$&$-0.405\pm0.069$&$0.177\pm0.013$&0.693\\
5.207&$0.00224\pm0.00011$&$-15.205\pm0.061$&$-0.534\pm0.065$&$0.104\pm0.008$&0.637\\
6.562&$0.00339\pm0.00030$&$-14.497\pm0.074$&$-1.021\pm0.069$&$0.094\pm0.011$&0.934\\
 \hline
Total & ...                                 & ...                            & ...                         & $5.234 \pm 0.327$ & ... \\ 
 \hline
\end{tabular}
\end{center}
\caption{\small Table of Schechter Fitting Parameters for current work for $FUV$ data split by $FUV-r$ colour. The final row is the flux density for the composite population of all the above $FUV-r$ subsets.}
\label{FUVparams}
\end{table*}

\begin{table*}\tiny
\begin{center}
\begin{tabular}{lcccccc}
  \hline
 $NUV-r$ & $\phi^*$ & $M^*-5 \log_{10}h$ & $\alpha$ & $f_{\nu} \times 10^{25}$ & $\chi^{2} / \nu$ \\ 
 & ($h^{3}$Mpc$^{-3}$mag$^{-1}$) & (mag) &  & ($h$ ergs s$^{-1}$ Hz$^{-1}$ Mpc$^{-3}$) &  \\ 
  \hline
1.868&$0.00279\pm0.00039$&$-17.719\pm0.088$&$-1.495\pm0.051$&$2.608\pm0.472$&2.046\\
2.203&$0.00291\pm0.00026$&$-17.613\pm0.068$&$-1.158\pm0.047$&$1.574\pm0.181$&0.973\\
2.444&$0.00206\pm0.00017$&$-17.773\pm0.068$&$-1.120\pm0.045$&$1.248\pm0.132$&0.963\\
2.662&$0.00189\pm0.00014$&$-17.767\pm0.066$&$-0.976\pm0.049$&$1.036\pm0.105$&0.970\\
2.881&$0.00222\pm0.00011$&$-17.500\pm0.051$&$-0.688\pm0.057$&$0.862\pm0.062$&2.050\\
3.134&$0.00225\pm0.00010$&$-17.290\pm0.051$&$-0.571\pm0.055$&$0.715\pm0.046$&1.965\\
3.453&$0.00234\pm0.00010$&$-16.996\pm0.056$&$-0.477\pm0.060$&$0.568\pm0.038$&1.181\\
3.882&$0.00252\pm0.00009$&$-16.531\pm0.048$&$-0.377\pm0.059$&$0.402\pm0.023$&1.742\\
4.475&$0.00258\pm0.00013$&$-16.235\pm0.054$&$-0.723\pm0.047$&$0.315\pm0.024$&0.887\\
5.548&$0.00828\pm0.00023$&$-15.189\pm0.034$&$-0.500\pm0.034$&$0.379\pm0.015$&0.808\\
 \hline
 Total & ...                                 & ...                            & ...                         & $9.706 \pm 0.541$ & ... \\ 
\hline
\end{tabular}
\end{center}
\caption{\small Table of Schechter Fitting Parameters for current work for $NUV$ data split by $NUV-r$ colour. The final row is the flux density for the composite population of all the above $NUV-r$ subsets.}
\label{NUVparams}
\end{table*}

\subsection{Dust Corrections}

All the flux density numbers presented so far are uncorrected for dust attenuation within the host galaxy (Galactic extinction has already been accounted for). Much work has been invested into the attenuation caused by dust, and the effect has been investigated in detail in recent papers by \citet{Tuffs2004,Driver2007,Driver2008}. The implied typical (so not taking into account specific galaxy geometry, but the mean effect of dust for an integrated population) correction for the $FUV$ is reported to be a factor of 4.32, and for the $NUV$ a factor of 4.11, as taken from \citet{Driver2008}. This would suggest the unobscured flux density for our sample is  $f_{\nu-FUV}=22.61 \pm 1.41 $ $\times 10^{25} h$ ergs s$^{-1}$ Hz$^{-1}$ Mpc$^{-3}$, $f_{\nu-NUV}=39.89 \pm 2.22 $ $\times 10^{25} h$ ergs s$^{-1}$ Hz$^{-1}$ Mpc$^{-3}$.

A better estimate can be made by only applying the \citet{Driver2008} corrections to the late-type selection, since we would typically expect little dust in our early-type sample. This gives a dust corrected $0.013 \le z \le 0.1$ late-type flux density of $f_{\nu-FUV}=22.13 \pm 1.41 $ $\times 10^{25} h$ ergs s$^{-1}$ Hz$^{-1}$ Mpc$^{-3}$, $f_{\nu-NUV}=38.10 \pm 2.22 $ $\times 10^{25} h$ ergs s$^{-1}$ Hz$^{-1}$ Mpc$^{-3}$. If these dust corrected values are added to the non-dust corrected early-type sample we can produce a final flux density for the Universe, with proper consideration made for the average dust attenuation. It is worth highlighting that the late-type galaxy sample dominates the integrated flux, thus the effect of only applying dust correction to the early-type galaxies changes the flux densities by only a few percent. Table \ref{finalflux} presents the final integrated flux densities for the local Universe for $FUV$ and $NUV$, and the implied SFR.

\begin{table*}\tiny
\begin{center}
\begin{tabular}{lccccccc}
  \hline
 & z-range & $f_{\nu} \times 10^{25}$ (original) & $f_{\nu} \times 10^{25}$ (corrected, Driver 2008) & SFR \\ 
 & z-range & $h$ ergs s$^{-1}$ Hz$^{-1}$ Mpc$^{-3}$ & $h$ ergs s$^{-1}$ Hz$^{-1}$ Mpc$^{-3}$ & $h$ M$_{\odot}$yr$^{-1}$Mpc$^{-3}$ \\ 
  \hline
 FUV composite & 0.013--0.1 & $ 5.234 \pm 0.327 \pm 0.660 $ & $ 22.24 \pm 1.41 \pm 2.80 $ & $ 0.0312 \pm 0.0020 \pm 0.0040 $\\ 
 FUV late & 0.013--0.1 & $ 5.123 \pm 0.327 \pm 0.646 $ & $ 22.13 \pm 1.41 \pm 2.79 $ & $ 0.0310 \pm 0.0020 \pm 0.0039 $ \\ 
 FUV early & 0.013--0.1 & $ 0.111 \pm 0.010 \pm 0.014 $ & ... & $ 0.00016 \pm 0.00001 \pm 0.00002 $ \\
 \hline
 NUV composite & 0.013--0.1 & $ 9.706 \pm 0.541 \pm 1.170 $ & $ 38.54 \pm 2.22 \pm 4.81 $ & ... \\ 
 NUV late & 0.013--0.1 & $ 9.270 \pm 0.540 \pm 1.169 $ & $ 38.10 \pm 2.22 \pm 4.80 $ & ... \\ 
 NUV early & 0.013--0.1 & $ 0.436 \pm 0.016 \pm 0.055 $ & ... & ... \\
 \hline\hline
 FUV Wyder (2005) & 0--0.1 & $ 5.17 \pm 0.98 \pm 1.78 $ & $ 22.33 \pm 4.23 \pm 7.70 $ & $ 0.0313 \pm 0.0059 \pm 0.0108 $ \\
 NUV Wyder (2005) & 0--0.1 & $ 7.52 \pm 2.16 \pm 2.59 $ & $ 30.91 \pm 6.21 \pm 10.65 $ & ...\\
 \hline
 FUV Budavari (2005) & 0.07--0.13 & $ 4.54 \pm 1.17 \pm 1.38 $ & $ 19.61 \pm 5.05 \pm 5.96 $ & $ 0.0275 \pm 0.0071 \pm 0.0083 $ \\
 NUV Budavari (2005) & 0.07--0.13 & $ 7.25 \pm 1.72 \pm 2.20 $ & $ 29.80 \pm 7.07 \pm 9.04 $ & ... \\
 \hline
 NUV Wyder (2007) & 0.01--0.25 & $ 8.83 \pm 0.61 \pm 0.32 $ & $ 36.29 \pm 2.51 \pm 1.31 $ & ... \\
 \hline 
\end{tabular}
\end{center}
\caption{\small  Table of dust corrected flux densities, and the derived \citet{Kennicutt1998} SFRs. The first part of the errors contain the covariance errors carried through from various sources, and the second part include a 12.6\% error expected due to Cosmic Variance (see Driver \& Robotham 2010) and a 0.36\% error in our area estimation (for this work), a 34.5\% CV error for Wyder 2005, 30.4\% CV error for Budavari 2005 and 3.6\% CV error for Wyder 2007. The CV error dominates over all other sources for all but Wyder 2007.}
\label{finalflux}
\end{table*}

\subsection{Low-z SFR Comparison}

The SFRs quoted in Table \ref{finalflux} are somewhat higher than typically reported for UV inferred values taken straight from the literature, but they are consistent with the H$_{\alpha}$ SFR of $0.029+0.008-0.005$  $h$ M$_{\odot}$yr$^{-1}$Mpc$^{-3}$ given by \citet{PerezGonzalez2003}. It is clear from this table that with data currently available we are dominated hugely in any estimate of the local SFR by cosmic variance. Since the light output in the GALEX UV bands is dominated almost solely by short lived stars (age less than 100 Myr) it is not sensible to probe over deeper baselines in $z$ where age resolution will be washed out. For measuring the cosmic star formation history (CSFH) this is a particular concern since we expect it to drop so steeply from $z=1$ to $z=0$ \citep{Hopkins2006}. The limit on measuring the SFR at the `present' day will always be the cosmic variance of the local Universe, a rough prediction from the Driver \& Robotham CV calculator suggests that within $z=0.08$ (the redshift limit out to 1 Gyr using $H_{0}=71$kms$^{-1}$/Mpc, $\Omega_{M}=0.27$ and $\Omega_{\Lambda}=0.73$) the local SFR cannot be constrained to better than 4.3\% even if we could measure the UV flux density in the whole sky, a survey area 50 times larger than that used for this work.

In an effort to consolidate the various low redshift measurements of the SFR we conducted an up-to-date literature search, normalising the quoted SFRs to our $H_{0}=100$kms$^{-1}$, $\Omega_{M}=0.3$ and $\Omega_{\Lambda}=0.7$ cosmology. Further to this, where UV SFRs were quoted, the \citet{Driver2008} dust corrections were applied to the data in order to treat the data as consistently as possible. Where possible, the volume of the various surveys were determined based on quoted area coverage and redshift range, allowing us to calculate the implied CV of each value using the same Driver \& Robotham (2010) equation as discussed above. If sharp redshift boundaries were not used, then the median redshift of the survey was used to determine an approximate volume. These additional CV error estimates were added in quadrature to the random errors stated for each SFR, allowing us to compare how unexpected any deviations in SFRs truly are.

\begin{figure*}
\centerline{
\mbox{\includegraphics[width=3in]{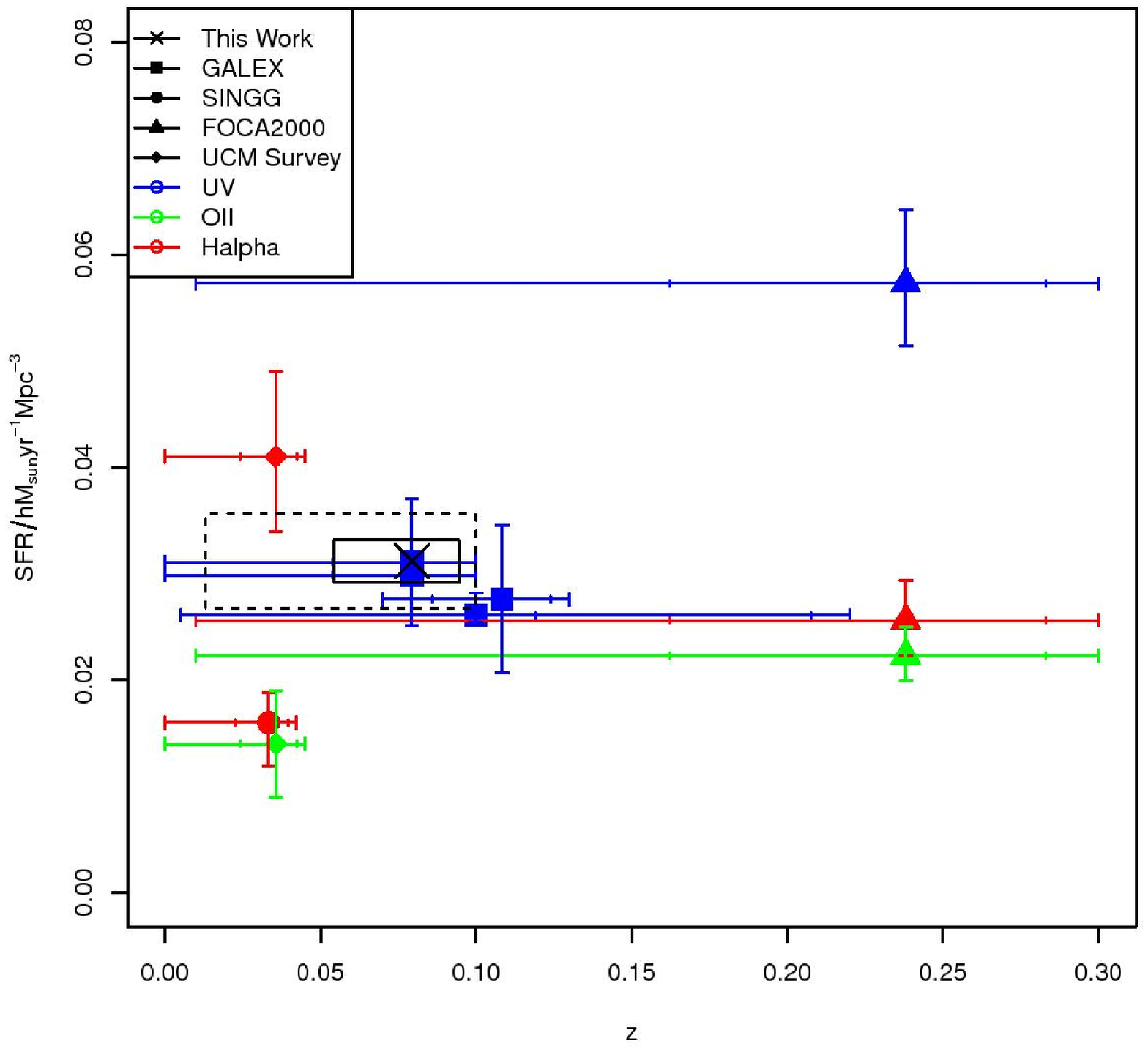}}
\mbox{\includegraphics[width=3in]{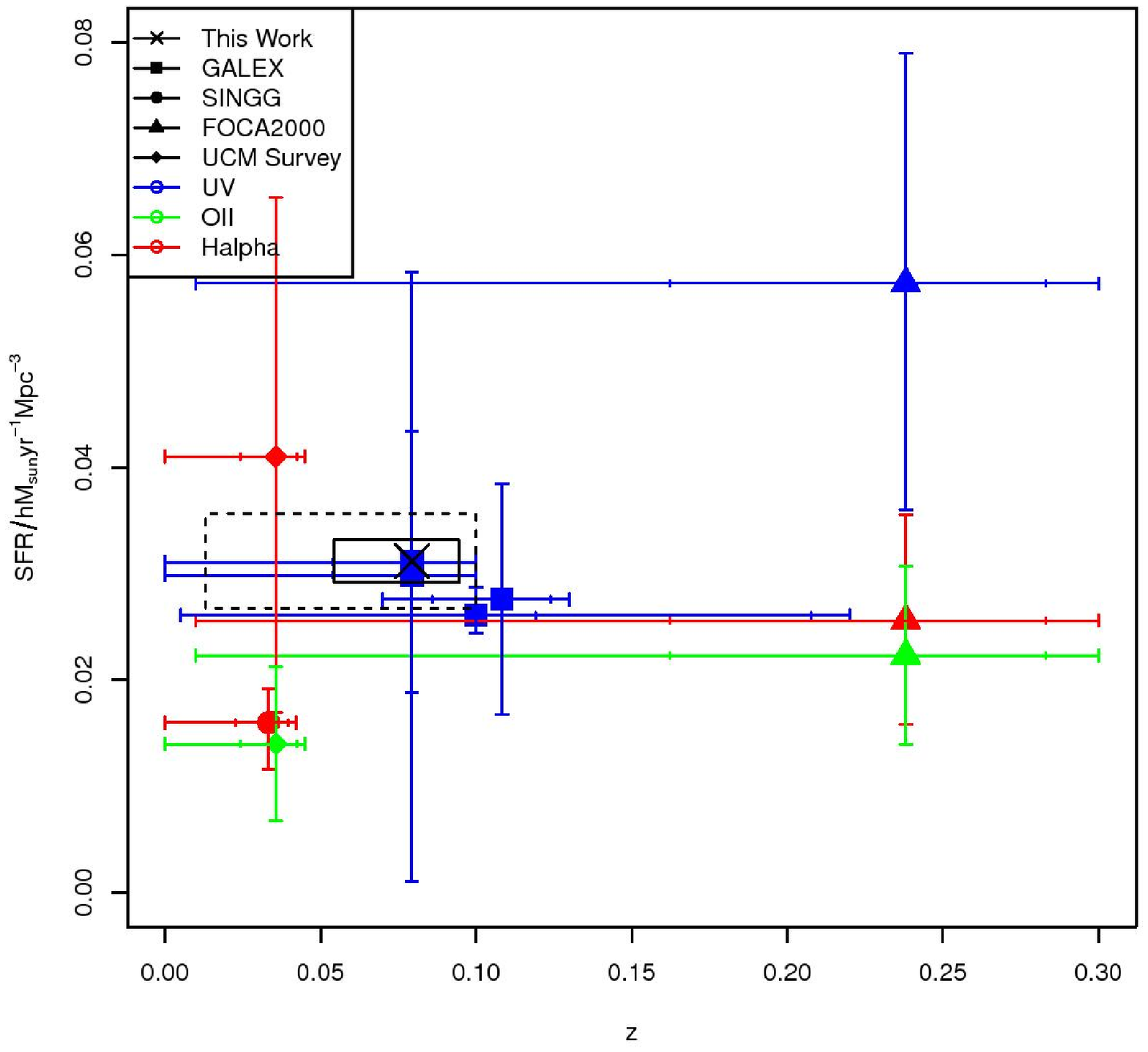}}
}
\caption{\small  Comparison of low redshift SFR indicators taken from the literature when factoring purely random errors (left side plot) or considering CV also (right side plot). Colour shows indicator type (UV, OII or H$\alpha$), and point style shows indicator source. Black boxes represent the results of this work (small error bars are the solid line, large error bars are the dotted line) On the x-axis, small error bars show the redshift at which 15\% and 85\% of the survey volume is reached, larger error bars show the extremities of the survey z-range. On the y-axis, error bars show the random errors quoted in the literature for the left side plot, error bars show the error implied when the worst case scenario CV error is added in quadrature for the right side plot. Exact values can be found in Table \ref{SFRtab}.}
\label{SFRcomp}
\end{figure*}

Table \ref{SFRtab} presents the relevant figures for these different SFR measurements, along with our new CV values. The table also explains what assumptions were made to obtain the SFR, i.e.\ what type of dust correction was applied (Driver, Calzetti or non-specific Balmer decrement correction) and what conversion to SFR was used: \citet{Kennicutt1998} or {\bf PEGASE} \citep{Fioc1997}. Figure \ref{SFRcomp} compares the SFRs as a function of redshift, separating out the quoted random errors found in the literature (small error bars) to the larger errors implied by adding the worst case scenario CV in quadrature (large error bars). The sources for this compendium of low-z SFRs are \citet{Budavari2005}, \citet{Gallego2002} \citet{Hanish2006}, \citet{PerezGonzalez2003}, \citet{Salim2007}, \citet{Schiminovich2005}, \citet{Sullivan2000} and \citet{Wyder2005}. These derive SFRs from utilising data from the following surveys: 2dFGRS, FOCA \citep{Milliard1992}, GALEX, HIPASS \citep{Meyer2004}, SDSS, SINGG \citep{Hanish2006} and the Universidad Complutense de Madrid Survey \citep[UCMS][]{Zamorano1994}.

Figure \ref{SFRcomp} suggests that once CV is taken into account most of the values for the local SFR found in the literature agree well with the new, much more tightly defined, SFR presented in this work. It should be highlighted that the CV quoted is the worst case scenario CV, calculated by converting the quoted volumes to cubes. Any geometry different to this will lower the CV, and actual values will often be a few percent smaller than stated here (see Driver \& Robotham 2010 for a detailed discussion of these effects). Another subtle effect is that some of the volumes used in various studies overlap to some degree, lowering their CV with respect to each other. Regardless, the CV is in practice always the dominant source of discrepancy, and the spread of values found in the literature can be explained by accounting for this source of error.

The lowest redshift SINGG derived H$\alpha$ and UCM derived OII SFRs do appear to be significantly low. Since these values probe significantly different redshift ranges (as indicated by the small error bars in Figure \ref{SFRcomp}), this could be interpreted as a real drop in cosmic SFR over the $\sim500$Myrs that covers the median redshift of our low redshift sample and median redshift of the two studies highlighted.

Perhaps the only significant difference between SFR indicators comes from using OII as opposed to UV or H$\alpha$. The OII SFR values are generally lower than all three indicators, and using the same surveys (UCMS and FOCA) they are lower than values recovered via H$\alpha$. For the UCMS derived SFRs the difference is well outside the random errors quoted (the appropriate errors to compare since the two measures share the same volume and suffer CV in the same direction).

\begin{table*}\tiny
\begin{center}
\begin{tabular}{lccccccccc}
 \hline
 & $z$-range & SFR & Area & Volume $\times10^{3}$ & CV & Indicator & Dust & SFR Source & Survey Source \\ 
 & &  $h$ M$_{\odot}$yr$^{-1}$Mpc$^{-3}$ & $\circ^{2}$ & ($h^{3}$Mpc$^{-3}$) & \% & & & &  \\ 
 \hline
This work 			& $0.013$--$0.1$ & $0.0312 \pm 0.0020$ 		& 833.13 	& 2,121 (exact) 	& 12.6 	& UV		& Driver	& Kennicutt & GALEX-SDSS \\
Budavari 2005 		& $0.07$--$0.13$ 	& $0.0276 \pm 0.0069$		& 43.9	& 202	(exact) 	& 30.4 	& UV		& Driver	& Kennicutt & GALEX-SDSS \\
Gallego 2002 		& $0$--$0.045$ 	& $0.014-0.005+0.005$		& 471.4	& 114	(exact) 	& 37.6	& OII		& Implied Balmer & Kennicutt & UCMS-HIPASS \\
Hanish 2006 		& $0$--$0.042$ 	& $0.016-0.0041+0.0028$	& 23,504	& 4,629 (scaled) 	& 9.4	& Halpha	& Helmboldt & Kennicutt & SINGG \\
Perez-Gonzalez 2003	& $0$--$0.045$ 	& $0.041+0.008-0.007$		& 195.0	& 47.1 (scaled) & 56.2	& Halpha	& Implied Balmer & Kennicutt & UCMS \\
Salim 2007               & $0.005$--$0.22$ & $ 0.0261-0.0006+0.0021$      & 645        & 16,046 (exact)      & 6.1        & SED      & Charlot \& Fall & Bruzual \& Charlot & GALEX-SDSS \\
Schiminovich 2005 	& $0$--$0.1$ 	& $0.0299-0.0048+0.0009$	& 3.141	& 8.02 (exact) 		& 95.2 	& UV		& Driver	& Kennicutt & GALEX-VVDS \\
Sullivan 2000 		& $0.01$--$0.3$ 	& $0.0573-0.0058+0.0070$	& 2.2	& 131	(exact) 	& 35.8	& UV		& Calzetti	& Pegase	 & FOCA \\
Sullivan 2000 		& $0.01$--$0.3$ 	& $0.0256-0.0033+0.0038$	& 2.2	& 131	(exact) 	& 35.8	& Halpha	& Calzetti	& Kennicutt & FOCA \\
Sullivan 2000 		& $0.01$--$0.3$ 	& $0.0223-0.0024+0.0027$	& 2.2	& 131	(exact) 	& 35.8	& OII		& Calzetti	& Kennicutt & FOCA \\
Wyder 2005 		& $0$--$0.1$ 	& $0.0311\pm0.0060$		& 56.73	& 145	(exact) 	& 34.5 	& UV		& Driver	& Kennicutt & GALEX-2dFGRS \\
\hline
\end{tabular}
\end{center}
\caption{\small  Table comparing different SFR measures at $z \le 0.3$ taken from the literature. Survey source shows the main combination of surveys required to generate the SFR. For most studies the secondary source supplies redshifts to allow LF measurements, in the case of Gallego (2002) the secondary source (HIPASS) is used to calibrate the normalisation. \citet{Schiminovich2005} is effectively derived from the measurements of \citep{Wyder2005}, so should not be considered a fully independent measure of SFR.}
\label{SFRtab}
\end{table*}

\section{Conclusions}

In this paper we have a calculated a number of flux densities for both dust-corrected and non-dust-corrected by integrating out to infinity various Schechter luminosity functions. We find the data are much better represented by dividing the populations into intense star forming populations and less active galaxies via their $NUV-r$ colour. This possesses a very strong bimodality, and allows dust corrections to be applied more appropriately to just the population of galaxies that are rapidly producing new stars and by inference will have the most dust attenuation.

Whilst flux densities, and the SFR they imply, have been calculated for a variety of stages of analysis, the final local SFR found for the low redshift $0.013 \le z \le 0.1$ Universe is $0.0312\pm0.0045$ $h$ M$_{\odot}$yr$^{-1}$Mpc$^{-3}$. The majority of the quoted uncertainty is due to cosmic variance (often called sample variance), which will dominate any calculation of the local SFR since it is strongly volume limited.

Various other low redshift SFRs were obtained from the literature and compared to values found in this work. Once CV is properly accounted for the values generally agree within errors, with the possible exception of the lowest redshift values--- these appear to be significantly on the low side. Also, the OII SFR indicator appears to predict generally lower values for the local SFR.

\section*{Acknowledgments}

ASGR acknowledges STFC funding for the GAMA post-doctoral fellowship. This work made use of the public data services provided by MAST and SDSS, the later funded by the Alfred P. Sloan foundation. Particular thanks goes to Bernie Shiao working at MAST for help in accessing the large volume of MIS data required for this work.

\newpage

\section*{Appendix}

\begin{table*}
\begin{center}
\begin{tabular}{lllllll}
M$_{\rm FUV}$ & Early $\phi$ ($h^{3}$Mpc$^{-3}$mag$^{-1}$) & Late $\phi$ ($h^{3}$Mpc$^{-3}$mag$^{-1}$) & All $\phi$ ($h^{3}$Mpc$^{-3}$mag$^{-1}$) \\
\hline
-20.25 & 0.000000 $\pm$ 0.000000 & 0.000000 $\pm$ 0.000000 & 0.000000 $\pm$ 0.000000 \\
-20.00 & 0.000000 $\pm$ 0.000000 & 0.000002 $\pm$ 0.000002 & 0.000002 $\pm$ 0.000002 \\
-19.75 & 0.000000 $\pm$ 0.000000 & 0.000000 $\pm$ 0.000000 & 0.000000 $\pm$ 0.000000 \\
-19.50 & 0.000000 $\pm$ 0.000000 & 0.000000 $\pm$ 0.000000 & 0.000000 $\pm$ 0.000000 \\
-19.25 & 0.000000 $\pm$ 0.000000 & 0.000003 $\pm$ 0.000002 & 0.000003 $\pm$ 0.000002 \\
-19.00 & 0.000000 $\pm$ 0.000000 & 0.000010 $\pm$ 0.000003 & 0.000010 $\pm$ 0.000003 \\
-18.75 & 0.000000 $\pm$ 0.000000 & 0.000025 $\pm$ 0.000004 & 0.000025 $\pm$ 0.000004 \\
-18.50 & 0.000000 $\pm$ 0.000000 & 0.000068 $\pm$ 0.000007 & 0.000068 $\pm$ 0.000007 \\
-18.25 & 0.000000 $\pm$ 0.000000 & 0.000144 $\pm$ 0.000009 & 0.000144 $\pm$ 0.000009 \\
-18.00 & 0.000000 $\pm$ 0.000000 & 0.000299 $\pm$ 0.000014 & 0.000299 $\pm$ 0.000014 \\
-17.75 & 0.000000 $\pm$ 0.000000 & 0.000461 $\pm$ 0.000018 & 0.000461 $\pm$ 0.000018 \\
-17.50 & 0.000001 $\pm$ 0.000001 & 0.000756 $\pm$ 0.000024 & 0.000757 $\pm$ 0.000024 \\
-17.25 & 0.000000 $\pm$ 0.000001 & 0.001055 $\pm$ 0.000030 & 0.001055 $\pm$ 0.000030 \\
-17.00 & 0.000002 $\pm$ 0.000001 & 0.001415 $\pm$ 0.000040 & 0.001417 $\pm$ 0.000040 \\
-16.75 & 0.000004 $\pm$ 0.000002 & 0.001866 $\pm$ 0.000055 & 0.001870 $\pm$ 0.000055 \\
-16.50 & 0.000009 $\pm$ 0.000003 & 0.002130 $\pm$ 0.000070 & 0.002139 $\pm$ 0.000070 \\
-16.25 & 0.000026 $\pm$ 0.000004 & 0.002722 $\pm$ 0.000096 & 0.002748 $\pm$ 0.000097 \\
-16.00 & 0.000037 $\pm$ 0.000005 & 0.003004 $\pm$ 0.000120 & 0.003041 $\pm$ 0.000120 \\
-15.75 & 0.000052 $\pm$ 0.000006 & 0.003369 $\pm$ 0.000158 & 0.003421 $\pm$ 0.000158 \\
-15.50 & 0.000098 $\pm$ 0.000008 & 0.004063 $\pm$ 0.000213 & 0.004160 $\pm$ 0.000213 \\
-15.25 & 0.000135 $\pm$ 0.000010 & 0.003953 $\pm$ 0.000233 & 0.004087 $\pm$ 0.000233 \\
-15.00 & 0.000191 $\pm$ 0.000012 & 0.003752 $\pm$ 0.000276 & 0.003943 $\pm$ 0.000277 \\
-14.75 & 0.000267 $\pm$ 0.000016 & 0.004223 $\pm$ 0.000375 & 0.004489 $\pm$ 0.000375 \\
-14.50 & 0.000297 $\pm$ 0.000018 & 0.004848 $\pm$ 0.000526 & 0.005145 $\pm$ 0.000526 \\
-14.25 & 0.000377 $\pm$ 0.000025 & 0.004869 $\pm$ 0.000610 & 0.005246 $\pm$ 0.000610 \\
-14.00 & 0.000396 $\pm$ 0.000029 & 0.004099 $\pm$ 0.000753 & 0.004495 $\pm$ 0.000754 \\
\end{tabular}
\end{center}
\label{appenFUV}
\caption{Summed galaxy densities for $FUV$. Absolute magnitudes are for the bin centres. The columns show the summed densities for early, late and all galaxies.}
\end{table*}

\begin{table*}
\begin{center}
\begin{tabular}{lllllll}
M$_{\rm NUV}$ & Early $\phi$ ($h^{3}$Mpc$^{-3}$mag$^{-1}$) & Late $\phi$ ($h^{3}$Mpc$^{-3}$mag$^{-1}$) & All $\phi$ ($h^{3}$Mpc$^{-3}$mag$^{-1}$) \\
\hline
-20.25 & 0.000000 $\pm$ 0.000000 & 0.000002 $\pm$ 0.000002 & 0.000002 $\pm$ 0.000002 \\
-20.00 & 0.000000 $\pm$ 0.000000 & 0.000000 $\pm$ 0.000000 & 0.000000 $\pm$ 0.000000 \\
-19.75 & 0.000000 $\pm$ 0.000000 & 0.000001 $\pm$ 0.000001 & 0.000001 $\pm$ 0.000001 \\
-19.50 & 0.000000 $\pm$ 0.000000 & 0.000014 $\pm$ 0.000003 & 0.000014 $\pm$ 0.000003 \\
-19.25 & 0.000000 $\pm$ 0.000000 & 0.000042 $\pm$ 0.000005 & 0.000042 $\pm$ 0.000005 \\
-19.00 & 0.000000 $\pm$ 0.000000 & 0.000093 $\pm$ 0.000008 & 0.000093 $\pm$ 0.000008 \\
-18.75 & 0.000000 $\pm$ 0.000000 & 0.000211 $\pm$ 0.000012 & 0.000211 $\pm$ 0.000012 \\
-18.50 & 0.000000 $\pm$ 0.000000 & 0.000352 $\pm$ 0.000015 & 0.000352 $\pm$ 0.000015 \\
-18.25 & 0.000000 $\pm$ 0.000001 & 0.000603 $\pm$ 0.000020 & 0.000604 $\pm$ 0.000020 \\
-18.00 & 0.000001 $\pm$ 0.000001 & 0.000885 $\pm$ 0.000025 & 0.000886 $\pm$ 0.000025 \\
-17.75 & 0.000003 $\pm$ 0.000002 & 0.001281 $\pm$ 0.000032 & 0.001285 $\pm$ 0.000032 \\
-17.50 & 0.000011 $\pm$ 0.000003 & 0.001597 $\pm$ 0.000040 & 0.001608 $\pm$ 0.000040 \\
-17.25 & 0.000020 $\pm$ 0.000004 & 0.002046 $\pm$ 0.000053 & 0.002066 $\pm$ 0.000053 \\
-17.00 & 0.000038 $\pm$ 0.000005 & 0.002339 $\pm$ 0.000066 & 0.002377 $\pm$ 0.000067 \\
-16.75 & 0.000098 $\pm$ 0.000008 & 0.002877 $\pm$ 0.000088 & 0.002976 $\pm$ 0.000089 \\
-16.50 & 0.000152 $\pm$ 0.000010 & 0.003304 $\pm$ 0.000114 & 0.003456 $\pm$ 0.000115 \\
-16.25 & 0.000258 $\pm$ 0.000013 & 0.003572 $\pm$ 0.000138 & 0.003831 $\pm$ 0.000138 \\
-16.00 & 0.000372 $\pm$ 0.000016 & 0.003875 $\pm$ 0.000177 & 0.004248 $\pm$ 0.000178 \\
-15.75 & 0.000491 $\pm$ 0.000018 & 0.004879 $\pm$ 0.000245 & 0.005370 $\pm$ 0.000246 \\
-15.50 & 0.000606 $\pm$ 0.000021 & 0.003988 $\pm$ 0.000253 & 0.004594 $\pm$ 0.000254 \\
-15.25 & 0.000699 $\pm$ 0.000024 & 0.004829 $\pm$ 0.000341 & 0.005528 $\pm$ 0.000342 \\
-15.00 & 0.000841 $\pm$ 0.000030 & 0.005162 $\pm$ 0.000442 & 0.006004 $\pm$ 0.000443 \\
-14.75 & 0.000867 $\pm$ 0.000035 & 0.005247 $\pm$ 0.000536 & 0.006113 $\pm$ 0.000537 \\
-14.50 & 0.000865 $\pm$ 0.000041 & 0.004634 $\pm$ 0.000697 & 0.005499 $\pm$ 0.000698 \\
-14.25 & 0.000876 $\pm$ 0.000047 & 0.004561 $\pm$ 0.000853 & 0.005437 $\pm$ 0.000855 \\
-14.00 & 0.000924 $\pm$ 0.000057 & 0.005962 $\pm$ 0.002068 & 0.006886 $\pm$ 0.002069 \\
\end{tabular}
\end{center}
\label{appenNUV}
\caption{Summed galaxy densities for $NUV$. Absolute magnitudes are for the bin centres. The columns show the summed densities for early, late and all galaxies.}
\end{table*}

\end{document}